\documentclass[twocolumn,showpacs,showkeys,preprintnumbers,amsmath,amssymb]{revtex4}
\usepackage{epsfig,latexsym}
\usepackage{graphicx}
\usepackage{dcolumn}
\usepackage{bm}
\def\Caption[#1][#2][#3]{\caption{\label{#1}\small{{\bf #2.} #3}}}
\def\ddt[#1]{\frac{d#1(t)}{dt}}
\def\partialddt[#1]{\frac{\partial#1}{\partial t}}
\def\formula[#1][#2]{\begin{equation}\label{#1}#2\end{equation}}
\def\eq[#1]{equation~\ref{#1}}
\def\Eq[#1]{Equation~\ref{#1}}
\def\eqPair[#1][#2]{equations~\ref{#1} and \ref{#2}}
\def\EqPair[#1][#2]{Equations~\ref{#1} and \ref{#2}}
\def\eqs[#1][#2]{equation~\ref{#1} to \ref{#2}}
\def\Eqs[#1][#2]{Equation~\ref{#1} to \ref{#2}}
\def\cfeq[#1]{(cf. equation~\ref{#1})}
\def\sec[#1]{section~\ref{#1}}
\def\Sec[#1]{Section~\ref{#1}}
\def\cfsec[#1]{(cf. section~\ref{#1})}
\def\fig[#1]{figure~\ref{#1}}
\def\Fig[#1]{Figure~\ref{#1}}
\def\cffig[#1]{(cf. figure~\ref{#1})}
\def\bs{$\!\!$}
\def\img[#1]{{\scriptsize {\bf #1}}}
\def\size[#1]{(size #1$\times$#1 pixels)}
\def\Lena{{\scriptsize {\bf Lena}} }
\def\Peppers{{\scriptsize {\bf Peppers}} }
\def\Lap{\nabla^2}
\def\heaviside{H}
\def\masyx{max-diffusion layer}
\def\misyx{min-diffusion layer}

\def\masy{{\masyx} }
\def\misy{{\misyx} }

\def\dnex{dynamic normalization}
\def\Dnex{Dynamic normalization}
\def\dne{{\dnex} }
\def\Dne{{\Dnex} }
\def\aemx{adaptation-by-entropy-maximisation}
\def\Aemx{Adaptation-by-entropy-maximisation}
\def\aem{{\aemx} }

\def\pseudox{pseudo}

\def\Pseudox{Pseudo}

\def\KEIL{\mathcal K}
\def\KEILIAN[#1]{\KEIL_{\!#1}}
\def\keilian{\KEILIAN[\lambda]}
\def\poskeil{\KEIL_{+\infty}}
\def\negkeil{\KEIL_{-\infty}}
\def\pics{./}
\begin{document}


\title[Local to Global Normalization]{Local to Global Normalization Dynamic by Nonlinear Local Interactions}


\author{Matthias S. Keil}%
\email{matsATcvc.uab.es; threequarksATyahoo.com; AT=@}
\affiliation{%
Basic Psychology Department,
Faculty for Psychology,
University of Barcelona (UB),
Passeig de la Vall d'Hebron 171,
E-08035 Barcelona (Spain)}%
%
%
%
\date{\today}
\begin{abstract}
Here, I present a novel method for normalizing a finite set of numbers,
which is studied by the domain of biological vision.
Normalizing in this context means searching the maximum and minimum
number in a set and then rescaling all numbers such that they fit into
a numerical interval.  My method computes the minimum and maximum number
by two pseudo-diffusion processes in separate diffusion layers.  Activity of
these layers feed into a third layer for performing the rescaling operation.
The dynamic of the network is richer than merely performing a rescaling
of its input, and reveals phenomena like contrast detection, contrast
enhancement, and a transient compression of the numerical range of the
input.  Apart from presenting computer simulations, some properties of
the diffusion operators and the network are analyzed mathematically.
Furthermore, a method is proposed for to freeze the model's state
when adaptation is observed.
\end{abstract}
%
%
\pacs{84.35.+i, 87.18.Bb, 87.18.Hf, 87.18.Sn, 87.19.Dd, 89.75.Kd}
%
%
\keywords{Adaptation, normalization, diffusion, network}
%
\maketitle
\begin{figure*}
	\scalebox{0.6}{\includegraphics{\pics/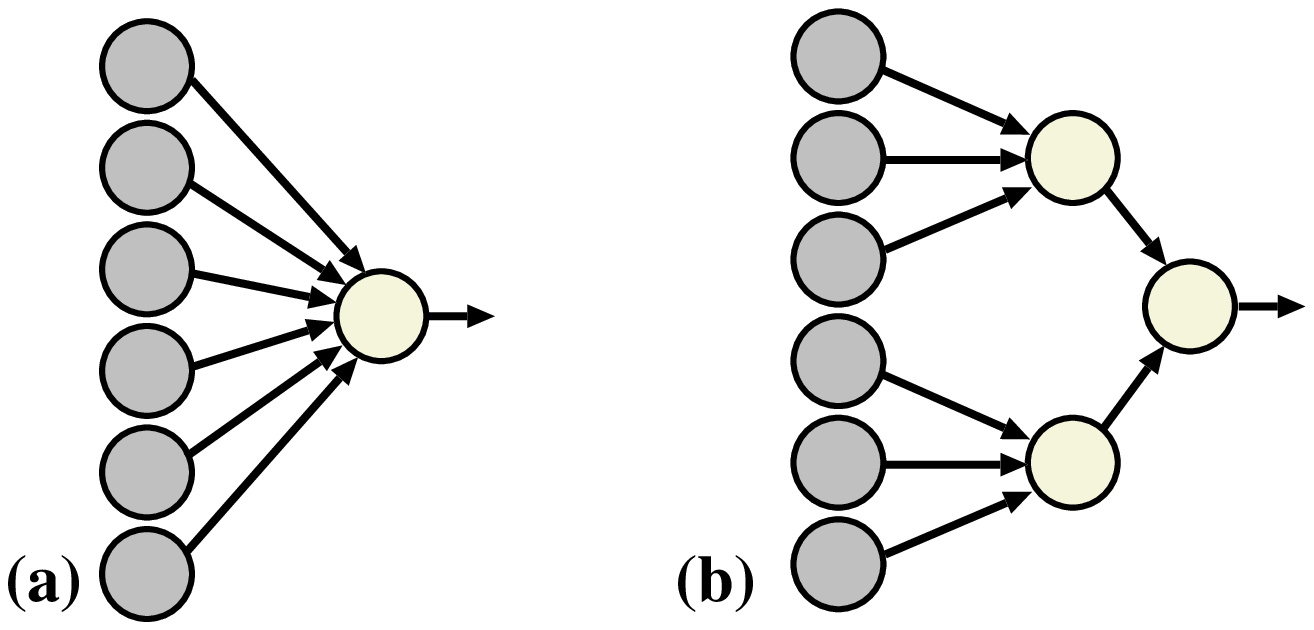}}
	\Caption[MinmaxNetworks][Possible network structures for extracting minimum
	or maximum activities of cells][{Each of the networks shown in this figure
	are supposed to select the maximum (or minimum) activity value
	among the leftmost units.  The selected maximum (or minimum) is
	available at the rightmost unit.  {\bf (a)} Two-layer network with global
	connectivity pattern.   {\bf (b)} Three-layer network which extracts
	in its second layer the (local) maximum of the units to which it is
	connected.  The rightmost unit subsequently selects from these local
	maximum the global one.}]
\end{figure*}
\section{Introduction}
%
%
What is the difference between adaptation and normalization? Are these just
two distinct processes, or can they be related? The purpose of this paper
is to develop a model whose dynamic smoothly proceeds from local adaptation
to global normalization.  Mathematical properties of the model are analyzed,
and its dynamical properties are evaluated with luminance images.  I study
the model within the framework of biological vision, where emphasis is laid
on understanding the emergence of adaptation within the model's dynamic.
Finally, a method is proposed for freezing the dynamic at the moment when
adaptation occurs.  But to begin with, I briefly describe how adaptation
and normalization contribute to information processing in the brain.\\
Adaptation refers to the adjustment of a sense organ to the intensity or
quality of stimulation \cite{Adaptation}.  There is agreement that adaptation 
is important for the function of nervous systems, since without corresponding
mechanisms any given neuron with its limited dynamic range would stay silent
or operate in saturation most of the time \cite{WalravenEtAl90}. 
When considering a population of \emph{cells} (e.g. formal processing units
or biological neurons), then adaptation is usually understood as a locally
acting process, which can be carried out independently for individual cells
or groups of cells, respectively (e.g., individual photoreceptors
\cite{CarpenterGrossberg,vanHateren2005,MatsJordi06} vs. groups of
photoreceptors\cite{GrossBrajovic03,GrossHong04}).  Thus, adaptation
refers to sensitivity adjustment of output signals as a function of
input signals.\\
Normalization on the other hand usually refers to establishing
standardized conditions for one or more qualities.  For example,
at some stage in the brain, the retinal image may have been normalized
with respect to illumination conditions, such that each face or object
is represented to have similar illumination patterns, and subsequent
recognition stages work in a more robust fashion.  Or, once a face
image has been detected by an artificial face recognition system,
it can be normalized with respect to head tilt or head rotation.
In this way a standardized candidate face is obtained, which
facilitates matching it to other standardized faces from a
database.\\
Normalization is also used for describing the establishment of standardized
conditions for a population of neurons.  In this context, normalization
processes usually act as gain control mechanisms.
For instance, Grossberg \cite{Grossberg83} proposed ``shunting competitive
networks'' (in his terms) for accurate signal processing in the presence
of noise to avoid the noise-saturation dilemma.  Because neurons have a
fixed input range, weak signals get masked by noise, and neurons' signal
only the noisy fluctuations in the input signal.  On the other hand, strong
signals cause neurons to saturate, and any variations within the input
cannot be distinguished.  Shunting networks implement the multiplicative
relationship between membrane voltages of neurons and conductance changes
that are caused by network input on the one hand and signals on the other.
This multiplicative relationship acts as a gain control mechanism that
enables these networks to automatically re-tune their sensitivity in
response to fluctuating background inputs.  As Grossberg demonstrated
\cite{Grossberg83}, such networks exhibit a normalization property
in the sense that the total (or pooled) activity of all neurons is
independent of the number of neurons.
Along these lines, Heeger and co-workers proposed a normalization model
to account for the observed non-linearities with the cortical simple cell
responses, such as response saturation and cross-orientation inhibition
\cite{CarandiniHeegerMovshon97,CarHee94,Heeger92}.
Similar to Grossberg's ``shunting competitive networks'', in Heeger's model
a neuron's output activity is adjusted by the pooled activity of a population
of many other neurons (``normalization pool'').  This ``normalization pool''
exerts divisive inhibition on the response of a target neuron, and in this
way it acts as a gain control mechanism for that cell.\\
The circuit models proposed by Grossberg and Heeger describe how responses
of a group of neurons can be normalized.  Both methods rely on the interaction
of some target neuron with a number of surrounding neurons.  The interaction is
brought about by hard-wiring the target neurons with surrounding neurons.  In
contrast, the normalization scheme introduced in this paper is based on diffusion
mechanisms, and thus interactions only take place between adjacent cells.
Specifically, within the scope of the present paper, normalization is understood
as mapping a set of numbers with finite but in principle arbitrary numerical
range onto a fixed target range (below we will see that non-trivial features
like contrast enhancement and adaptation phenomena emerge from a network which
implements this normalization mechanism).\\
Whereas in Grossberg's scheme the normalization process renders
the total activity of a group of cells independent of the number of cells
(\cite{Grossberg83}), with my definition of normalization it is clear that
in most cases the activity summed over all cells will depend on their number.
A further difference concerns the implementation of activity bounds.
In Grossberg's scheme, reversal potentials establish an upper (lower) bound
on the activity of each cell which can be reached by excitation (inhibition).
However, the highest activity value of the normalized cell population depends
on the activity of all other cells (as the total activity is constant).
In other words, one cannot rely on the presence of distinguished activity
values as it is the case in my approach.  In a normalized population of
my approach there is always at least one cell which has zero activity, and
at least one cell with activity one.\\
The usual proceeding for normalizing a set of numbers can be subdivided into two
successive stages.  First, the maximum and the minimum members of the set are
determined.  These two values are then used in a second stage for re-scaling
all set elements such that after re-scaling the elements fall into a pre-defined
numerical interval (or numerical range).\\
%
If we wish to design a corresponding algorithm for the first stage
of the just described process (i.e. finding the maximum and the minimum),
we would have to employ two memories for storing the \emph{current}
(i.e., a local) maximum and minimum, and compare these values successively
with all remaining set elements.  After we finished with comparing, the
memory would contain the \emph{global} maximum and minimum.  Because
every set member has to interact explicitly with the memories, the
whole process is said to involve global operations.  The global nature
is mirrored in the connection structure of a correspondingly designed
network.  \Fig[MinmaxNetworks](a)
shows a schematic drawing of such a network, where one distinguished
network unit shares connections with all the others.  This unit
is supposed to represent the maximum (or minimum) activity value of
the set of units to which it is connected to.  Due to its global
connectivity, however, our network seems not to be a very plausible
candidate for a ``biologically'' model, because (biological) neurons
are known to interact in a more local fashion.  This implausibility
can be relaxed by proposing an alternative connectivity pattern
(\fig[MinmaxNetworks](b)).\\
Nevertheless, the two units representing the maximum and the minimum,
respectively, need to interact subsequently again with the input units,
in order to put into effect the re-scaling operation that implements
the gain control mechanism.  This means that one would require yet another
set of non-local connections, analogously to the pattern shown in
\fig[MinmaxNetworks].
This led me to the question whether such normalization can be achieved
in a more ``biological'' or local fashion, or even by employing only
interactions between adjacent network units.  Presuming the existence
of corresponding mechanisms, one has to explore in addition whether
they could, in principle, be carried out by nerve cells in a biophysically
plausible way.\\
Below I present a network (the \emph{dynamical normalization network}), which
achieves normalization by means of lateral propagation of activity between
adjacent network cells.  To this end, parameterized diffusion operators were
developed.  In their limit cases, these operators implement non-linear and
non-conservative diffusion processes (``\pseudox-diffusion'').  The dynamic
of \pseudox-diffusion proceeds from local to global in a continuous
fashion, without utilizing any connectivity structure apart from coupling
among nearest neighbors.\\
The dynamic normalization network consists of a total of four layers: an input layer,
two diffusion layers, and the normalization or output layer, where all layers
interact.  Numerical simulations with luminance images revealed that the dynamic
of the normalization layer is functionally more rich than just performing a
re-scaling of its input.  Initially, the dynamic reveals contrast enhancement
similar to high-pass filtering.\\
Furthermore, under certain conditions, an adaptation phenomenon (``dynamic
compression'') can be observed in the initial phase of the dynamic.  As it is
described in detail below (section \ref{section_DynamicCompression}), the strength
of the dynamic compression effect depends on the size of high activity regions in
the input, and their relative positions with respect to other local maxima.
%
\section{Formal Description of Nonlinear Diffusion} 
%
The dynamic normalization network is based on nonlinear diffusion operators.
In order to proof some of their properties, it is necessary that the nonlinear
diffusion operators are differentiable.  Accordingly, we define at first an
operator $\mathcal{T}_{\lambda}[\cdot]$ which is parameterized over $\lambda$ as
\formula[minmaxNonlinearity][{	\mathcal{T}_{\lambda}[x]=
				\frac{\eta x}{1+e^{-\lambda x}}}]
where $\eta$ is a normalization constant that is defined as
\formula[minmaxNormConstant][{\eta=1+e^{-|\lambda|}.}]
Through the specific choice of $\lambda$, we can ``steer'' the operator
$\mathcal{T}_{\lambda}[\cdot]$ continuously from linearity
($\mathcal{T}_{0} \equiv \mathcal{T}_{\lambda=0}$)
\formula[minmaxLambdaLinear][{\lambda=0 \leadsto  \mathcal{T}_{0}[x]=x }]
to half wave rectification (i.e. selection of the maximum between zero
and its argument)
\formula[minmaxLambdaPositive][{\lim_{\lambda\rightarrow +\infty}
				T_\lambda[x]\equiv\max(x,0)}]
or inverse half wave rectification (i.e. selection of the minimum between
zero and its argument)
\formula[minmaxLambdaNegative][{\lim_{\lambda\rightarrow -\infty}
				T_\lambda[x]\equiv\min(0,x).}]
Notice that the operator satisfies $T_{-\infty}[-x]=-T_{+\infty}[x]$.
%
\subsection{Spatially continuous nonlinear diffusion equation in one dimension}
%
\def\EXP{e^{-\lambda z}}
\def\PEXP{e^{-\lambda |z|}}
\def\PEXP{e^{-\lambda |z|}}
\def\PLIM{\lim_{\lambda\rightarrow +\infty}}
\def\NEXP{e^{\lambda |z|}}
With the operator $\mathcal{T}_{\lambda}[\cdot]$, one can define a general
diffusion scheme which contains heat-diffusion as a special case for
$\lambda=0$.  To this end, consider, without loss of generality, the
general form of a diffusion equation for a quantity $f(x,t)$ (here
referred to as ``activity'')
\formula[minmaxLinear1D][{	\partialddt[f(x)] = \frac{\partial}{\partial x}
				\left(
					D(x) \frac{\partial f}{\partial x}
				\right)}]
where $D(x)\geq 0$ is the diffusion coefficient.  If $D(x)$ depends on $x$,
then the last equation describes a nonlinear diffusion process, otherwise
ordinary heat diffusion.  Consequently, by applying the operator
$\mathcal{T}_{\lambda}[\cdot]$ on the gradients, the following
\emph{\pseudox-diffusion} process is obtained (which reduces to heat
diffusion for $\lambda=0$):
\formula[minmaxNonlinear1D][{	\partialddt[f(x)] = \frac{\partial}{\partial x}
				\left(
					D(x) \mathcal{T}_{\lambda} \left[
					\frac{\partial f}{\partial x} \right]
				\right).}]
By defining $z(x) \equiv \partial f(x) / \partial x$ and differencing we obtain
\formula[minmaxNonlinear1D_2][{	\partialddt[f(x)] = \frac{\partial D(x) }
				{\partial x} \mathcal{T}_{\lambda}[z]
				+  D(x) \frac{\partial \mathcal{T}_{\lambda}[z]}
				{\partial z} \frac{\partial z(x)}{\partial x}.}]
If $D(x)=D=\mathrm{const.}$, the last equation reduces to
\formula[minmaxNonlinear1D_3][{\partialddt[f(x)] = D
				\frac{\partial \mathcal{T}_{\lambda}[z]}
				{\partial z}
				\frac{\partial^2 f(x)}{\partial x^2}.}]
The last equation looks in fact like an ordinary diffusion equation if we
consider the factor $D\,\partial \mathcal{T}_{\lambda}[z] / \partial z$
as an ``effective diffusion coefficient''.  But which effect has the
derivative $\partial \mathcal{T}_{\lambda}[z] / \partial z$?  In appendix
\ref{ProofOne} it is shown that it approximates a Heaviside (or step)
function $\heaviside$ for $\PLIM$, that is
\formula[minmaxDerivative4lambda_oo_Heavi][{	\PLIM \frac{\partial
						\mathcal{T}_{\lambda}[z]}
						{\partial z}
						\approx \heaviside(z).}]
In analogy to the previous case it can be shown that
\formula[minmaxDerivative4lambda_noo_Heavi][{\lim_{\lambda\rightarrow -\infty}
	\frac{\partial \mathcal{T}_{\lambda}[z]}{\partial z} \approx 
	-\heaviside(z).}]
For a given cell, $\lambda$ specifies the ratio between negative and positive
influx into the cell from its neighbors.
Consider the case $\lambda \rightarrow \infty$ for a cell at position $x$.
If the activity of any adjacent cell is higher, then the gradient
$z(x) \equiv \partial f(x) / \partial x$ will be positive and an influx of
activity to cell $x$ takes place, because in \eq[minmaxNonlinear1D_2] the
\pseudox-diffusion term $\partial z(x)/\partial x$ is multiplied by one as a consequence
of \eq[minmaxDerivative4lambda_oo_Heavi].  \Eq[minmaxDerivative4lambda_oo_Heavi]
also implies that any negative gradient at $x$ will make the \pseudox-diffusion
term be multiplied by zero, and thus prevents an influx of negative activity into
cell $x$.  The essence of this mechanism is that activity at $x$ can only increase
until any gradient has dissipated.   As an alternative, this mechanism may be
understood as an auto-adaptive diffusion constant which regulates its value
according to the current gradient at $x$ (\fig[TwoCellSystem]).\\
For $\lambda \rightarrow -\infty$ the mechanism works just vice versa, and
the activity for a cell at position $x$ may only decrease.  The linear (or heat)
diffusion equation is obtained for $\lambda=0$, where both a positive-valued
and a negative-valued influx can enter the cell.
\begin{figure}[hbpt]
	  {\bf \large (a)}\scalebox{0.45}{\includegraphics{\pics/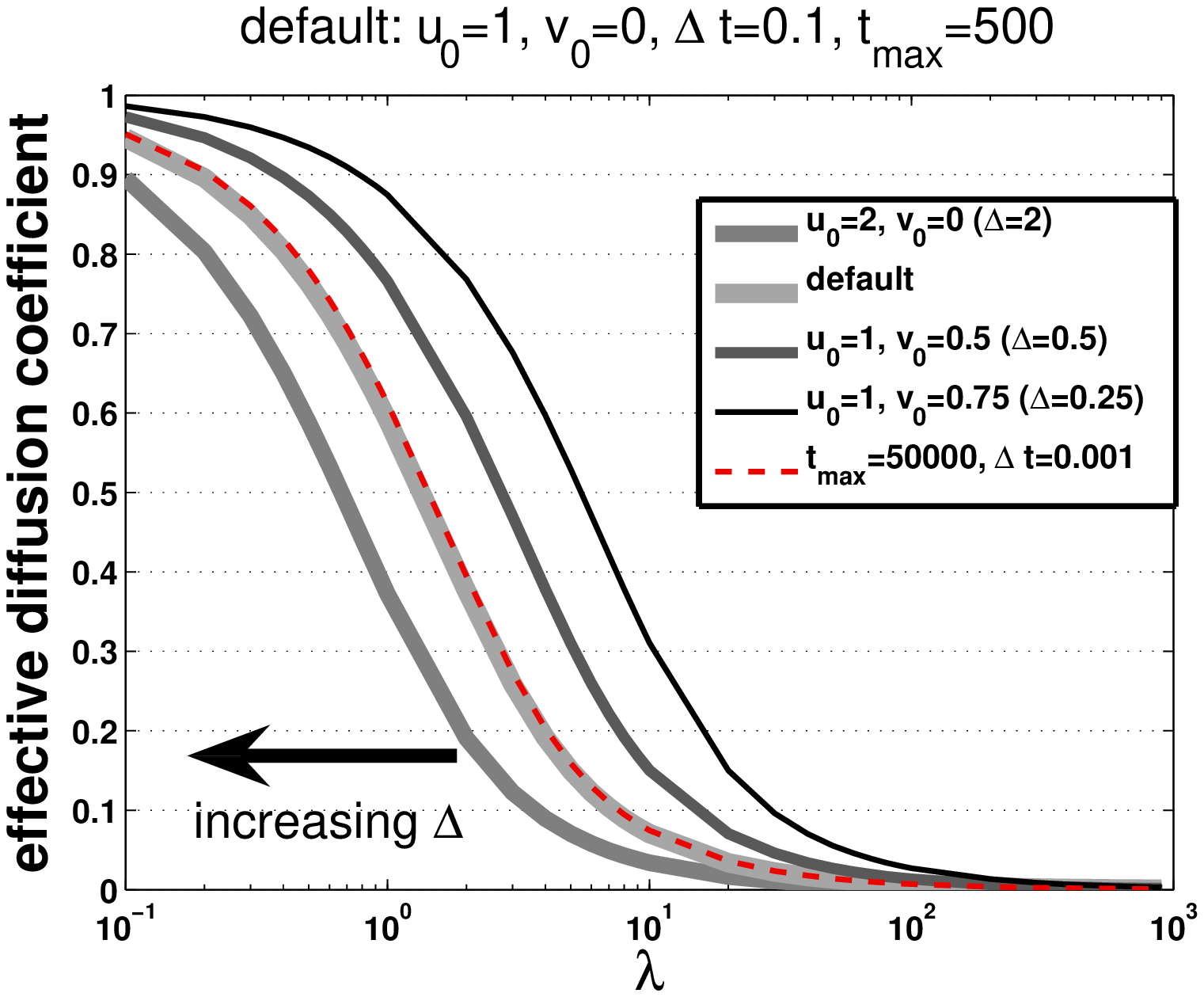}}
	  {\bf \large (b)}\scalebox{0.45}{\includegraphics{\pics/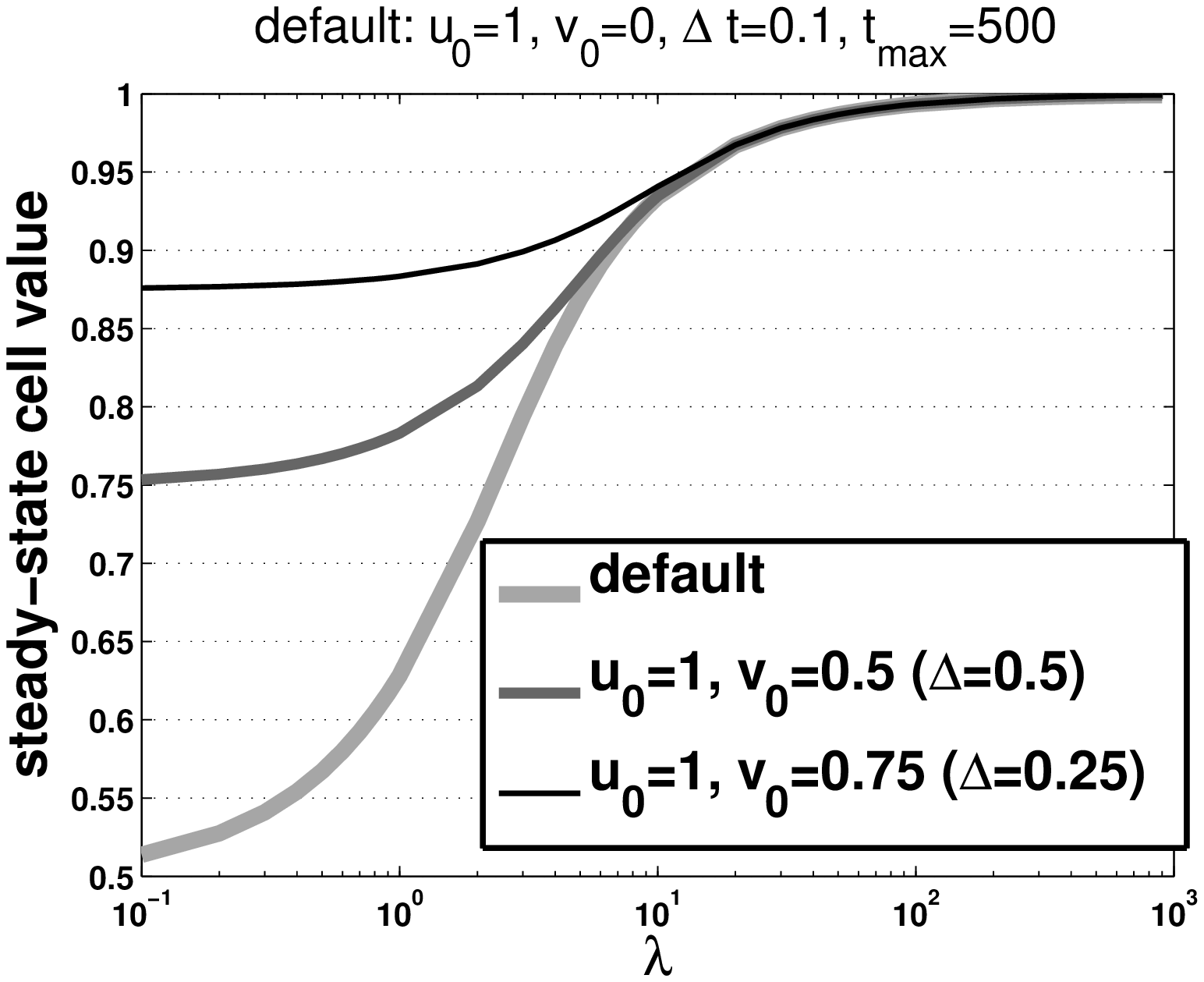}}
	\Caption[TwoCellSystem][Diffusion for $\mathbf{0 \leq \lambda<\infty}$][{%
	{\bf (a)} The plot relates $\lambda$ of \eq[twocells] (ordinate) to the
	effective diffusion constant $\gamma$ of \eq[twocells_surrogate]
	(abscissa).  The different curves relate to different simulation
	parameters as indicated in the legend.  Parameters that do not appear
	in the legend correspond to default values as indicated in the
	figure heading ($\Delta t=0.1$ integration step size, $t_\mathrm{max}=500$
	iteration limit).  An
	increase of the value of the gradient $\Delta\equiv u_0-v_0$ at
	$t=0$ makes the curves $\gamma(\lambda)$ shift to the left
	(arrow).  With $\Delta t=0.1$ or smaller, results do not depend
	in a significant way on integration step size
	(dashed line which overlaps with the curve for the default
	case).  With increasing increments $\Delta t$, however, all
	solid lines displace to the left by the same amount.
	{\bf (b)} The steady-state cell values $u_\infty \equiv v_\infty$
	as a function of $\lambda$ show a sigmoidal relationship which
	smoothly passes from heat diffusion ($u_\infty\equiv [u_0+v_0]/2$
	for $\lambda=0$) to implementing a maximum operation
	($u_\infty\equiv \max[u_0,v_0]$ for $\lambda=\infty$).}]
\end{figure}
\begin{figure*}
	\scalebox{0.75}{\includegraphics{\pics/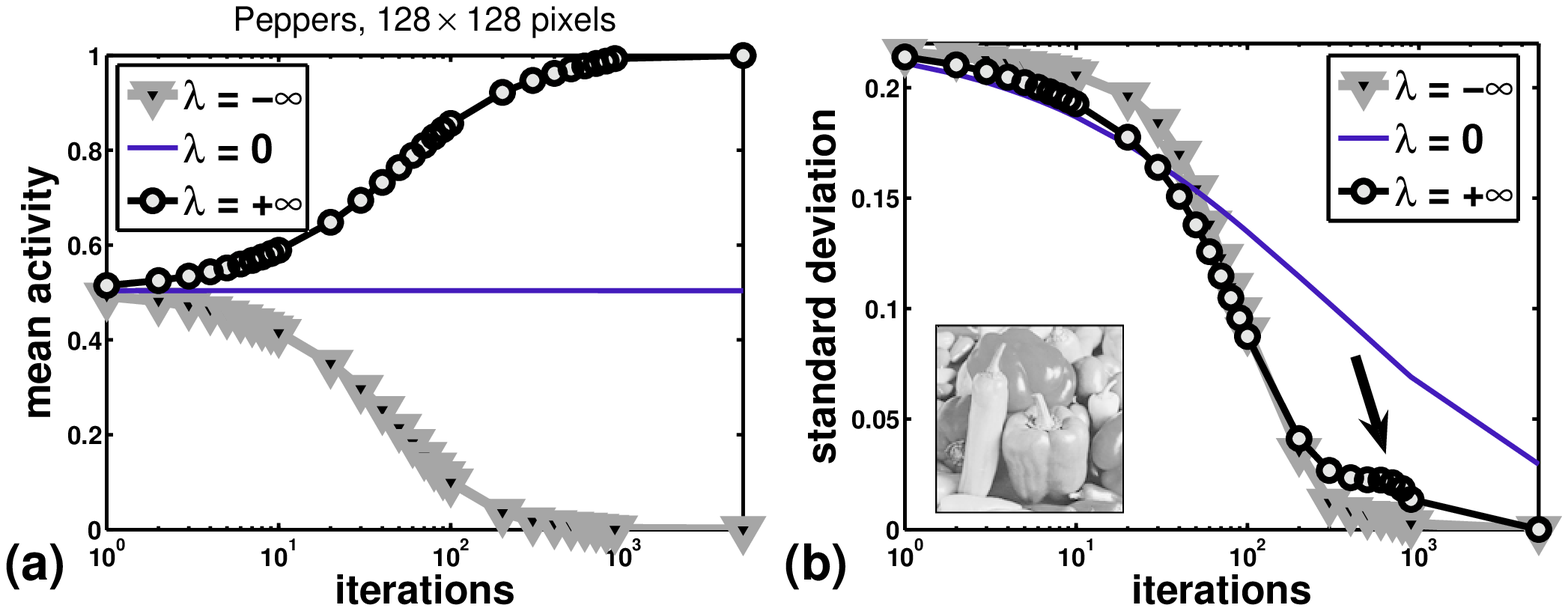}}
	\Caption[minmaxMeanStdvPeppers][\Pseudox-diffusion converges
	faster than heat diffusion (\Peppers image)][{Curves show the temporal
	evolution of the mean activity together with standard deviations
	for heat diffusion ``$\lambda=0$'' (no symbols, eq. \ref{minmaxDiffMean}),
	max-diffusion ``$\lambda=+\infty$'' (circles, eq. \ref{minmaxDiffMax}),
	and min-diffusion ``$\lambda=-\infty$ ''(triangles, eq. \ref{minmaxDiffMin}).
	For computing the mean activity,  averaging took place over all values of the
	respective (\pseudox-) diffusion layer.
	The \Peppers image ($0 \leq s_{ij} \leq 1$, size $128 \times 128$
	pixels, inset) defined the initial state of each layer. {\bf (a)}
	Mean activity remains constant with time with heat diffusion
	(heat diffusion is conservative), but approaches the minimum
	(maximum) value of $s$ in the \misy (\masyx).  {\bf (b)} The
	minimum (maximum) is finally adopted by all cells $a_{ij}$ ($b_{ij}$),
	as indicated by decreasing standard deviations.
	Compared to heat diffusion, the \pseudox-diffusion systems
	converge in fewer simulation time steps to an uniform state,
	but for their simulation more computations per time step
	are needed (cf. \eq[minmaxNonlinearity]).  Moreover, \pseudox-diffusion
	does not converge after $2\cdot 128$ iterations
	(the largest distance between two cells on a $N\times N$ grid is $2N$
	in a Manhattan architecture).  A single iteration is insufficient
	to propagate a maximum from one cell to the next, as all
	diffusion operators are normalized by the number of adjacent cells
	(in addition, $D\Delta t\leq1/2$, see \sec[Methods]).  Note
	further the that standard deviation of the max-diffusion layer
	has a local maximum (arrow).  This at first sight
	paradoxical effect is explained with \fig[minmaxMeanStdvGrayBlackWhite].}]
\end{figure*}
%
%
%
%
\subsection{Intermediate values of $\lambda$ for a two cell system}
%
Intermediate values of $\lambda$ attenuate either negative ($\lambda>0$) or
positive influx ($\lambda<0$).  The amount of attenuation depends on $\lambda$.
To illustrate, consider a simplified \pseudox-diffusion system which consists
only of two cells $u(t)$ and $v(t)$:
\begin{eqnarray}\label{twocells}
	\partialddt[u] & = & \mathcal{T}_{\lambda}[v-u]\\
	\partialddt[v] & = & \mathcal{T}_{\lambda}[u-v]\nonumber
\end{eqnarray}
Furthermore, we define the following \emph{surrogate} system
\begin{eqnarray}\label{twocells_surrogate}
	\partialddt[a] & = & \gamma (b-a)\\
	\partialddt[b] & = & a-b \nonumber
\end{eqnarray}
with a diffusion coefficient $\gamma$.  Note that because diffusion coefficients
are different for $a(t)$ and $b(t)$ (that is, $\gamma$ and $1$, respectively),
the last equation implements a nonlinear diffusion system.  Without loss of
generality, we assume $\lambda>0$, and $u_0-v_0>0$ at $t=0$.  Furthermore,
let both diffusion systems have the same initial conditions $u_0=a_0$ and
$b_0=v_0$.  With this configuration of parameters, the influx into cell
$u$ is negative, and will be attenuated because of $\lambda>0$.  Dependent
on the precise value of $\lambda$, the steady-state values of $u_\infty$
and $v_\infty$ will be situated somewhere between $(u_0+v_0)/2$ for $\lambda=0$,
or $\max(u_0,v_0)$ for $\lambda\rightarrow +\infty$.  Now, to understand
the behavior for $0<\lambda<\infty$, the diffusion coefficient $\gamma$
is (numerically) determined such that both diffusion systems (equations
\ref{twocells} and \ref{twocells_surrogate}) have the same equilibrium
state, that is $u_\infty = a_\infty$ and $v_\infty = b_\infty$ (and
also $u_\infty = v_\infty$).\\
With the assumptions $\lambda>0$ and $u_0-v_0>0$, it follows that $\gamma<1$,
because in order to obtain the same steady-state values for both diffusion
systems, the negative influx into cell $a$ needs to be attenuated.
\Fig[TwoCellSystem]\emph{a} shows that in this case the effective
diffusion coefficient $\gamma$ and $\lambda$ have a sigmoidal relationship.
The sigmoid shifts to the left as a function of $\Delta\equiv u_0-v_0$
(or equivalently $a_0-b_0$).\\
\Fig[TwoCellSystem]\emph{b} shows that steady-state values as a function
of $\lambda$ also follow a sigmoidal relationship.  Cell values at
convergence smoothly pass from heat diffusion
($u_\infty\equiv [u_0+v_0]/2$ for $\lambda=0$) to implementing
a maximum operation ($u_\infty\equiv \max[u_0,v_0]$ for $\lambda=\infty$).
Analogous considerations hold for negative values of $\lambda$.
%
\subsection{Spatially discrete \pseudox-diffusion equation in two dimensions}
%
%
Based on a centered finite difference representation of the Laplacian operator,
we define a parameterized diffusion operator acting on a function $f(x,y)$ as
\formula[minmaxKeilianDef][{
\begin{array}{rc}
	\keilian f(x,y)= 
		& \mathcal{T}_{\lambda}\left[f(x+1,y)-f(x,y)\right]	\\
	+	& \mathcal{T}_{\lambda}\left[f(x-1,y) - f(x,y)\right]	\\
	+	& \mathcal{T}_{\lambda}\left[f(x,y+1)-f(x,y)\right]	\\
	+	& \mathcal{T}_{\lambda}\left[f(x,y-1) - f(x,y)\right]	\\
\end{array}%
}]
where a grid spacing of $\Delta x = \Delta y = 1$ is assumed.  We will make
use of the following compact notation
\formula[minmaxCompactPos][{	\poskeil \equiv \lim_{\lambda\rightarrow +\infty}
				\keilian}]
and
\formula[minmaxCompactNeg][{	\negkeil \equiv \lim_{\lambda\rightarrow -\infty}
				\keilian }.]
Note that $\KEILIAN[\lambda=0] \approx \Lap$ from \eq[minmaxLambdaLinear].\\
In order to formulate a spatially discrete \pseudox-diffusion
scheme, we consider a  diffusion layer (i.e. a finite grid on which diffusion
takes place) with an equal number $N$ of rows $i$ and columns $j$, that is
$1 \leq i,j \leq N$.  We use a discrete-in-space and continuous-in-time
notation, where $f_{ij}=f(j,i)$, $f_{i,j+1}=f(j+1,i)$ and so on
\cite{rem_rowscols}.  With the above definitions, heat diffusion
is described as:
\formula[minmaxDiffMean][{\partialddt[f_{ij}] = D \cdot \KEILIAN[0] f_{ij}(t)}]
where $D=const.$ is the diffusion coefficient.  The process is assumed to
start at time $t=t_0$ with the initial condition $f_{ij}(t_0)=s_{ij}$.
From now on we assume that the $s_{ij}$ represent an intensity or luminance
distribution (i.e., ``$s$ represents a gray level image).
\begin{figure*}
	\scalebox{0.75}{\includegraphics{\pics/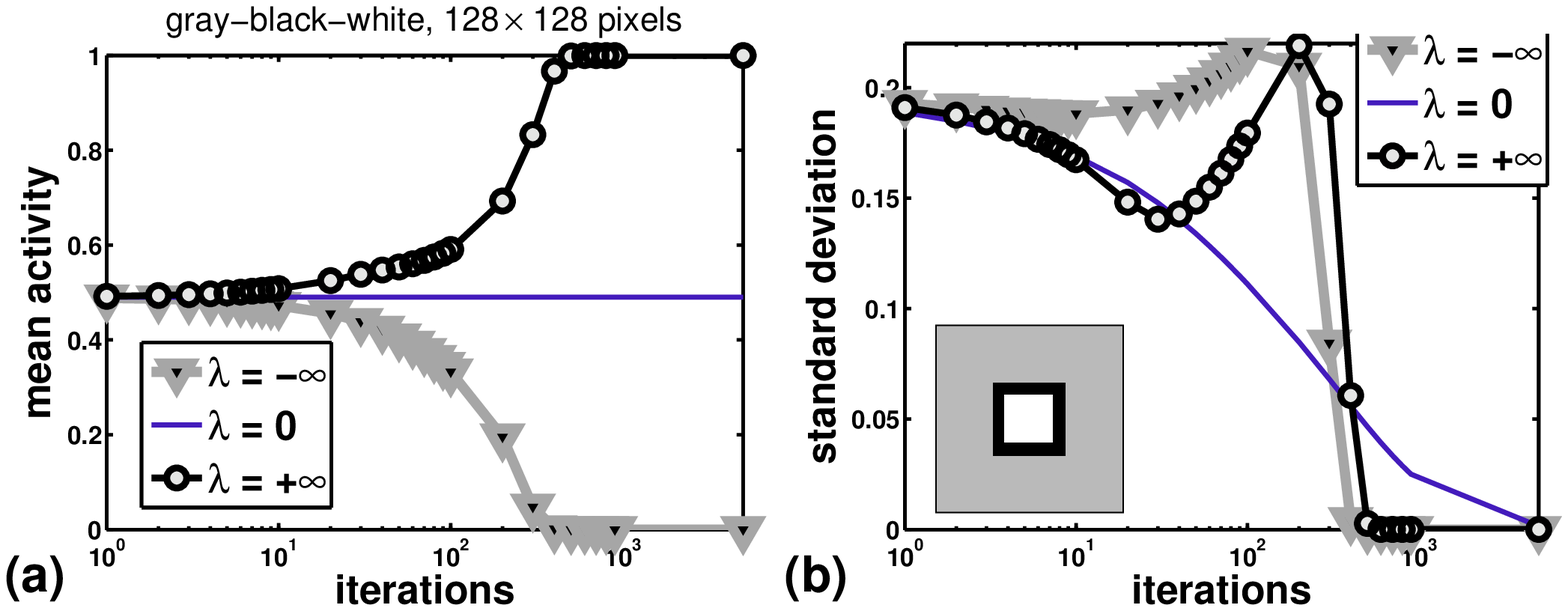}}
	\Caption[minmaxMeanStdvGrayBlackWhite][Why standard deviations
	can increase with \pseudox-diffusion][{Standard deviations of
	the \misy and the \masy reach a local or global maximum if at some
	time in the layers a configuration is obtained which consists
	of two domains with each domain having a different activity
	level (similar to a luminance step).  Of course, a necessary
	precondition is that the initially provided
	configuration (here three domains with gray $s_{ij}=0.5$, black
	$s_{ij}=0$, and white $s_{ij}=0$, see inset) has a smaller
	standard deviation than the ``step''-like configuration which
	is generated as an intermediate state.  In the \masyx, for
	example, a step-like configuration is reached as soon as the
	black frame is dissolved from both sides (i.e., ``eaten''
	by the gray and the white region).
	The effect depends on luminance levels and the relative
	amount of black, white, and gray.  It can be also obtained for
	different layouts of the regions (e.g., gray-white-black or black-gray-white). }]
\end{figure*}
Since diffusion takes place in a bounded domain (i.e. we have a finite number
$N\times N$ of grid points), and we also use adiabatic boundary conditions
(i.e. there is neither inward flux nor any flux outward over the domain boundary,
i.e. $\partial f_{ij}/\partial{t}=0$ for $(i,j)\in \{(i,0),(i,N),(0,j),(N,j)\}$)
\cite{rem_adiabatic}, the total activity described by \eq[minmaxDiffMean] does
not depend on time (\fig[minmaxMeanStdvPeppers]), that is
\formula[minmaxConstSum][{\sum_{i,j}^N f_{ij}(t)=\mathrm{const.}}]
The last expression expresses that diffusion is conservative -- activity
is neither created nor destroyed.  Although the 2-D heat diffusion equation
cannot create new activity levels which have not already been present at time
$t_0$ \cite{Koenderink84}, it can create extrema in activity domains that
have a dimension greater than one (cf. \cite{LifshitzPizer1990}, p.532).
A {\it \misy} will eventually compute the minimum of all values and
is defined as:
%
\formula[minmaxDiffMin][{	\partialddt[a_{ij}] = D \cdot \negkeil a_{ij}(t).}]
A {\it \masy} will eventually compute the maximum of all values and 
is defined as:
%
\formula[minmaxDiffMax][{	\partialddt[b_{ij}] = D \cdot \poskeil b_{ij}(t).}]
We assume equal initial conditions $a_{ij}(t_0)=b_{ij}(t_0)=s_{ij}$ for the \misy
and the \masy at time $t=t_0$.\\
Whereas \eq[minmaxDiffMean] preserves its total activity, the \misy and \masyx,
respectively, do not.  The total activity of the \misy decreases with time and
converges to (\fig[minmaxMeanStdvPeppers])
\formula[minmaxLower][{	\lim_{t \rightarrow \infty} \sum_{i,j}^N a_{ij}(t)
			= N^2 \min_{i,j}\{s_{ij}\}.}]
The total activity of the \masy increases with time and converges to
(\fig[minmaxMeanStdvPeppers])
\formula[minmaxUpper][{	\lim_{t \rightarrow \infty} \sum_{i,j}^N b_{ij}(t)
			= N^2 \max_{i,j}\{s_{ij}\}.}]
In other words, all cells $a_{ij}$ of the \misy will finally contain the
global minimum of the input $s_{ij}$
\begin{equation}\label{MINmax}
	A:=\min_{i,j}\{s_{ij}\} = \lim_{t \rightarrow \infty} a_{ij}(t)
					\ \ \forall i,j 
\end{equation}
and all cells $b_{ij}$ of the \masy will end up with the global maximum
\begin{equation}\label{minMAX}
	B:=\max_{i,j}\{s_{ij}\} = \lim_{t \rightarrow \infty} b_{ij}(t)
					\ \ \forall i,j .
\end{equation}
This can be explained as follows.  A cell $a_{ij}$ of the \misy may only
{\it decrease} its activity from one time step to the next, until any
activity gradient between $a_{ij}$ and its nearest neighbors has
dissipated.  As a consequence, $a_{ij}$ adopts the minimum activity
value of the neighborhood, including itself.
Because the last arguments apply to \emph{all} cells $a_{ij}$, eventually
all cells will adopt the minimum activity $\min_{i,j}\{s_{ij}\}$ at
convergence.  Convergence occurs if $a_{ij}=a_{kl}\,\forall\,(i,j),(k,l)$
(i.e. when no more activity gradient exists).  The dynamic of the
process is illustrated by \fig[minmaxPeppersDiff].\\
In an analogous way, in the diffusion process described by the \masyx,
all cells $b_{ij}$ could only {\it increase} their activity, given the
existence of any activity gradient.  If any cell has a maximum activity
value, then finally all cells will adopt this maximum, since only then
all gradients have dissipated.\\
Hence, both nonlinear diffusion systems are non-conservative, because
they do not fulfill requirements analogous to \eq[minmaxConstSum].
\begin{figure*}
	\scalebox{0.70}{\includegraphics{\pics/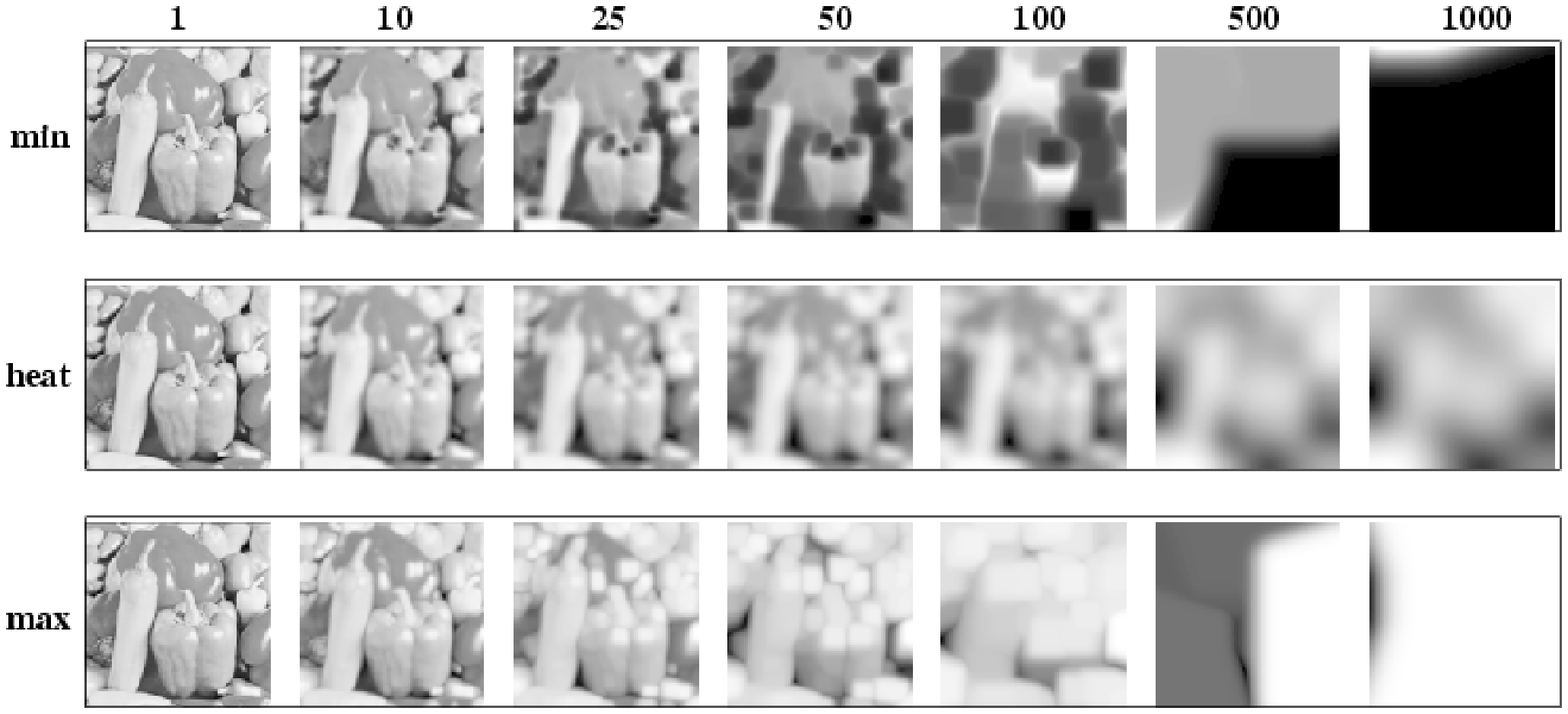}}
	\Caption[minmaxPeppersDiff][Snapshots of diffusion states][{Images
	show snapshots of max-diffusion ($a_{ij}$, first row), heat diffusion
	($f_{ij}$, middle row), and min-diffusion ($b_{ij}$, last row) for the
	\Peppers image \size[128].  The numbers indicate elapsed iterations.
	Whereas heat diffusion just blurs the image, min-diffusion and
	max-diffusion create ``islands'' corresponding to local minima
	and maxima, respectively.  With increasing time, islands decrease
	in number and increase in size, until eventually a single island
	occupies the whole region.  Then, the \misy and the \masy have
	eventually computed the global minimum and maximum, respectively.
	Due to our boundary conditions (see methods), diffusion is faster
	at domain boundaries (see \sec[Methods]).  Brighter gray levels
	correspond to higher cell activities.}]
\end{figure*}
%
%
\section{\label{dynamicNormalisation}%
		Dynamic normalization by next neighbor interactions} 
%
Equipped with the \pseudox-diffusion operators defined in the last section,
we are now ready to define the {\it dynamic normalization} network. The network
normalizes a given input $s_{ij}$ with respect to numerical range, but without
taking resort to any global memory for determining the minimum and maximum.
Rather, the global minimum and maximum are computed in the \misy and the \masyx,
respectively, by only exchanging information between adjacent cells.\\
We start with the following linear scaling scheme, which is typically used for
normalizing a fixed set of numbers (again, see introduction):
\formula[minmaxLinearScaling][	c_{ij}=\frac{s_{ij}-a_{ij}}{b_{ij}-a_{ij}}
				\ \ \mathrm{where}\ \ 1 \leq i,j \leq N.]
Because of \eq[MINmax] and \ref{minMAX} the variable $c_{ij}$ will contain
(after a sufficiently long time) a normalized representation of $s_{ij}$,
that is
\formula[minmaxMapping][{s_{ij} \in [A,B] \mapsto c_{ij} \in [0,1]}]
($A$ and $B$ are the global minimum and maximum, respectively,
of $\{s_{ij}\}$).  To arrive at a fully dynamical system, we formally
interpret \eq[minmaxLinearScaling] as the steady-state solution of
\formula[minmaxDynamicNorm][{\partialddt[c_{ij}]=b_{ij} (0-c_{ij}) - a_{ij} (1-c_{ij})
				+ s_{ij}}]
which shall be called {\it \dnex}.  Notice that by using \dne we
naturally avoid the singularity of \eq[minmaxLinearScaling] that
occurs for $b_{ij}=a_{ij}$.\\
\begin{figure*}
	\scalebox{0.75}{\includegraphics{\pics/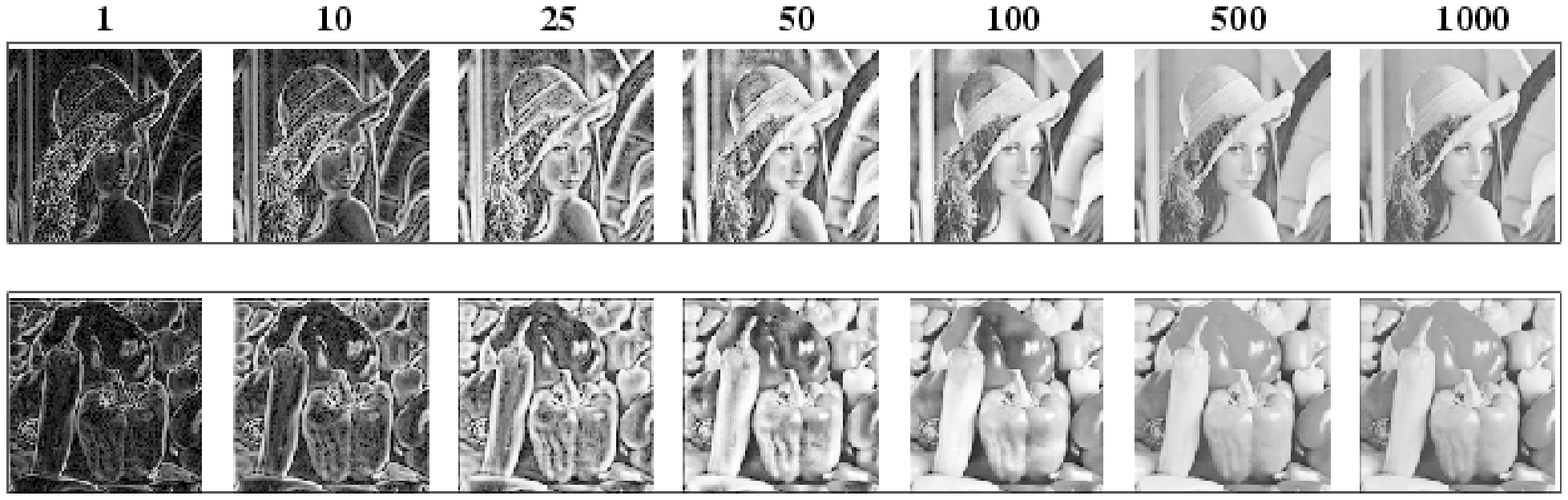}}
	\Caption[dneLenaPeppers][Snapshots of \dnex][{The images show
	\eq[minmaxDynamicNorm] at different time steps (indicated by
	the numbers) for images \size[128] \Lena (top row) and \Peppers
	(bottom row).  The dynamic begins similar to high-pass filtering
	(cf. \fig[spatfreqGrating]),
	proceeds with contrast enhancement and ``fills in'' image
	structures from their contrast contours, until finally
	a normalized version of the input image is obtained.
	Brighter gray levels indicate higher cell activities.
	In order to improve visualization, images were
	rescaled individually.}]
\end{figure*}
\begin{figure}
	\scalebox{0.5}{\includegraphics{\pics/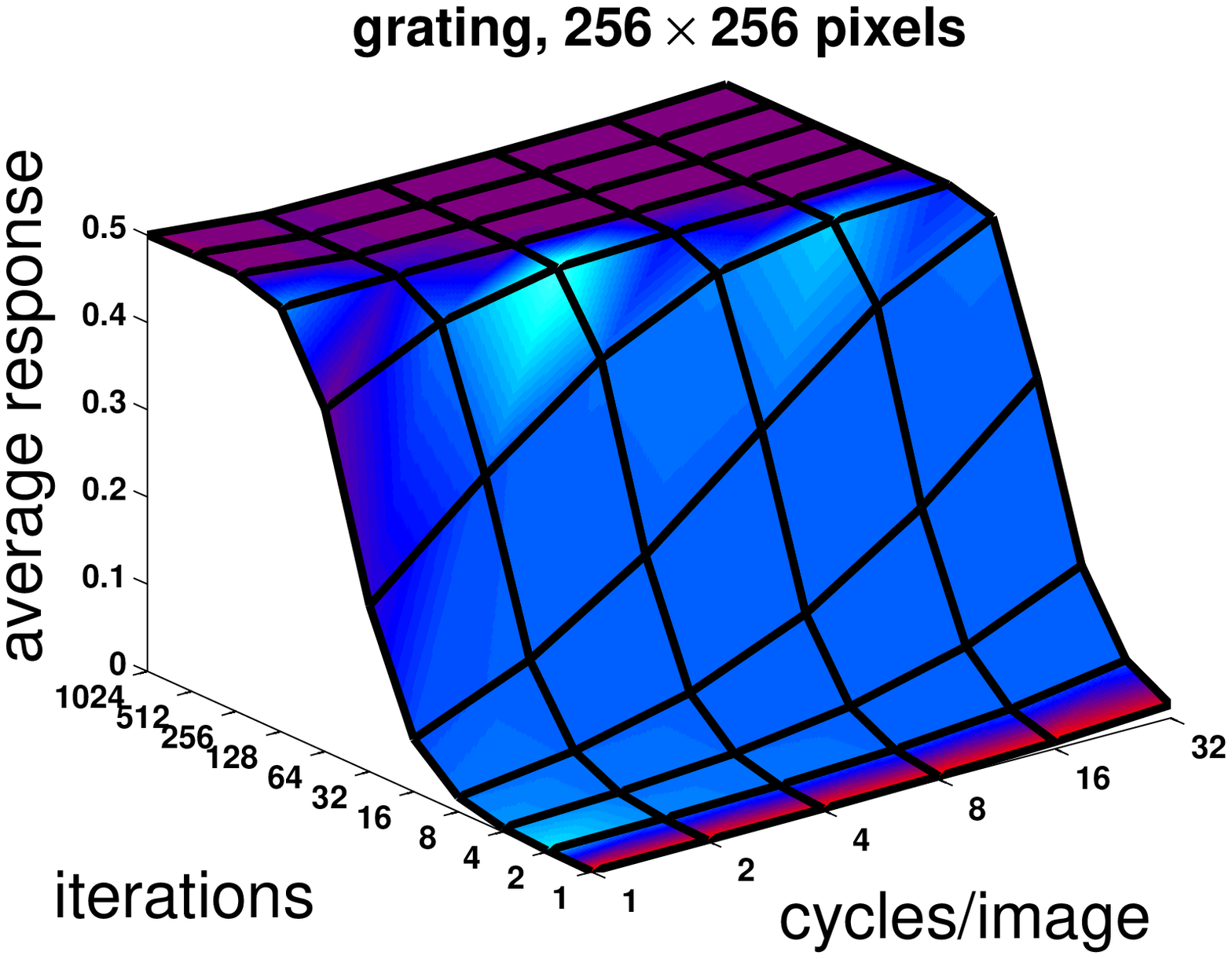}}
	\Caption[spatfreqGrating][Spatial frequency vs. time for a sine wave grating][{This experiment
	is analogous to \fig[minmaxChessContrast](b), but here for a sine wave grating \size[256].  At
	each time, the graphs show the maximum activity value of \eq[minmaxDynamicNorm].  Obviously,
	at a fixed number of iterations ($\lessapprox 64$), the \dne network's signal transmission
	characteristics is high-pass.  No frequency selectivity is observed after convergence
	($\gtrapprox 128$ iterations).}]
\end{figure}
\Fig[dneLenaPeppers] visualizes the state of \eq[minmaxDynamicNorm] at
different time steps.  Initially, the \dne process is similar
to high-pass filtering (\fig[spatfreqGrating]), what can
be explained as follows.  Contrasts are abrupt changes in luminance.
Consider a luminance change from dark to bright.
Then, the dark side has a local minimum, and the
bright side a local maximum, which propagates in the \misy
and the \masyx, respectively (\fig[minmaxPeppersDiff]).
When the local minimum (maximum)
has propagated to the position of the bright (dark) side
of the step, then the bright (dark) side will be normalized
to one (zero).  As the dynamic continues to evolve, local maxima and
minima propagate further, thereby ``eating'' (i.e., annihilating)
other smaller local maxima and minima.  In \fig[dneLenaPeppers],
this annihilation of local maxima and minima, respectively,
is visible through a gradual filling-in of image structures
from the boundaries.  A normalized version of the original
image is finally obtained when $\partial c_\mathrm{ij}/\partial t=0$.
Depending on \emph{(i)} how small the $s_\mathrm{ij}$ are, and \emph{(ii)} the
choice of integration step size $\Delta t$, the steady-state of $c_\mathrm{ij}$
can be reached with delay compared to the steady-states of $a_\mathrm{ij}$
and $b_\mathrm{ij}$, respectively.  This is now examined in more
detail.
%
\def\figscale{0.75} 
%
\begin{figure}
	  {\bf \large (a)}\scalebox{0.45}{\includegraphics{\pics/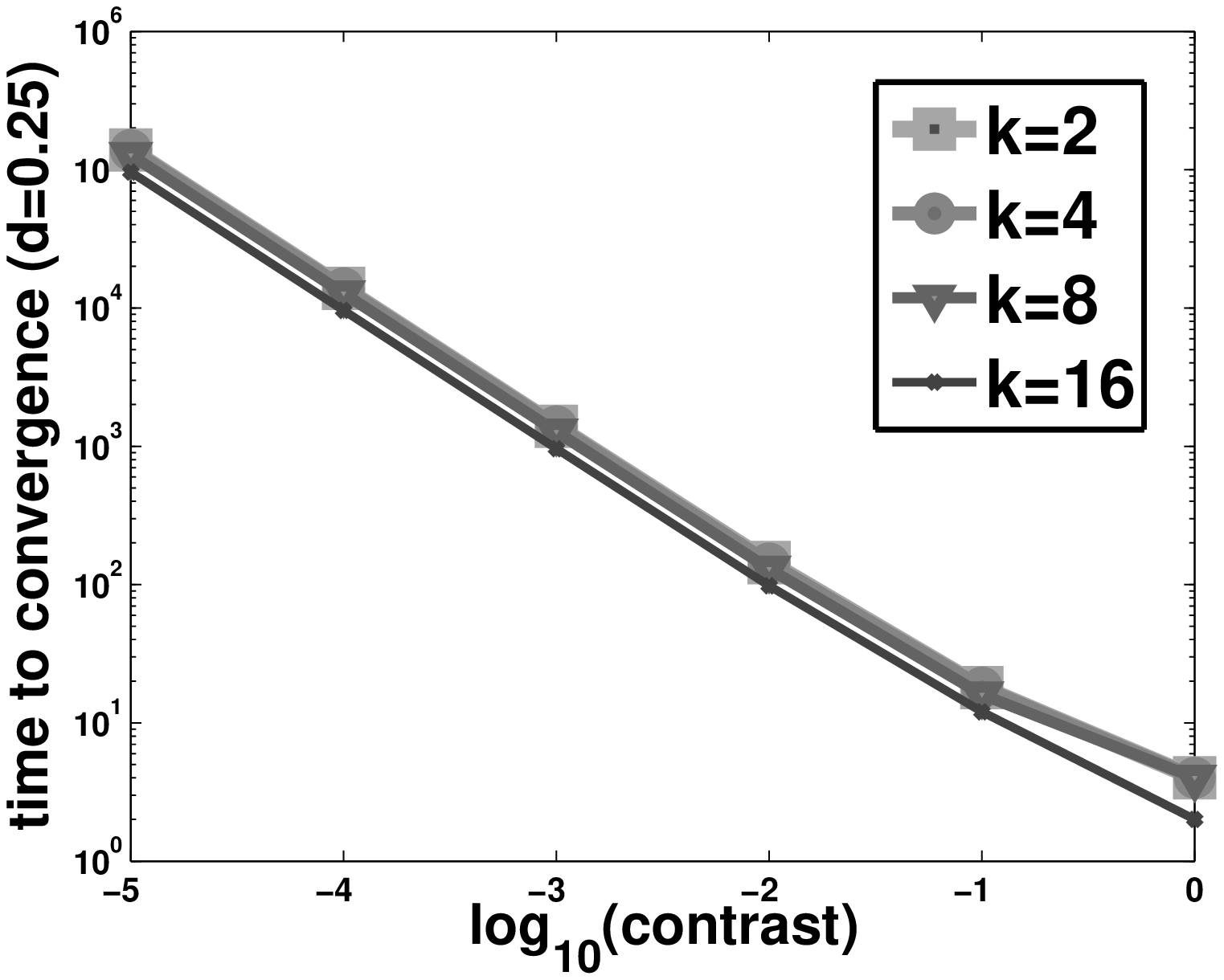}}
	  {\bf \large (b)}\scalebox{0.45}{\includegraphics{\pics/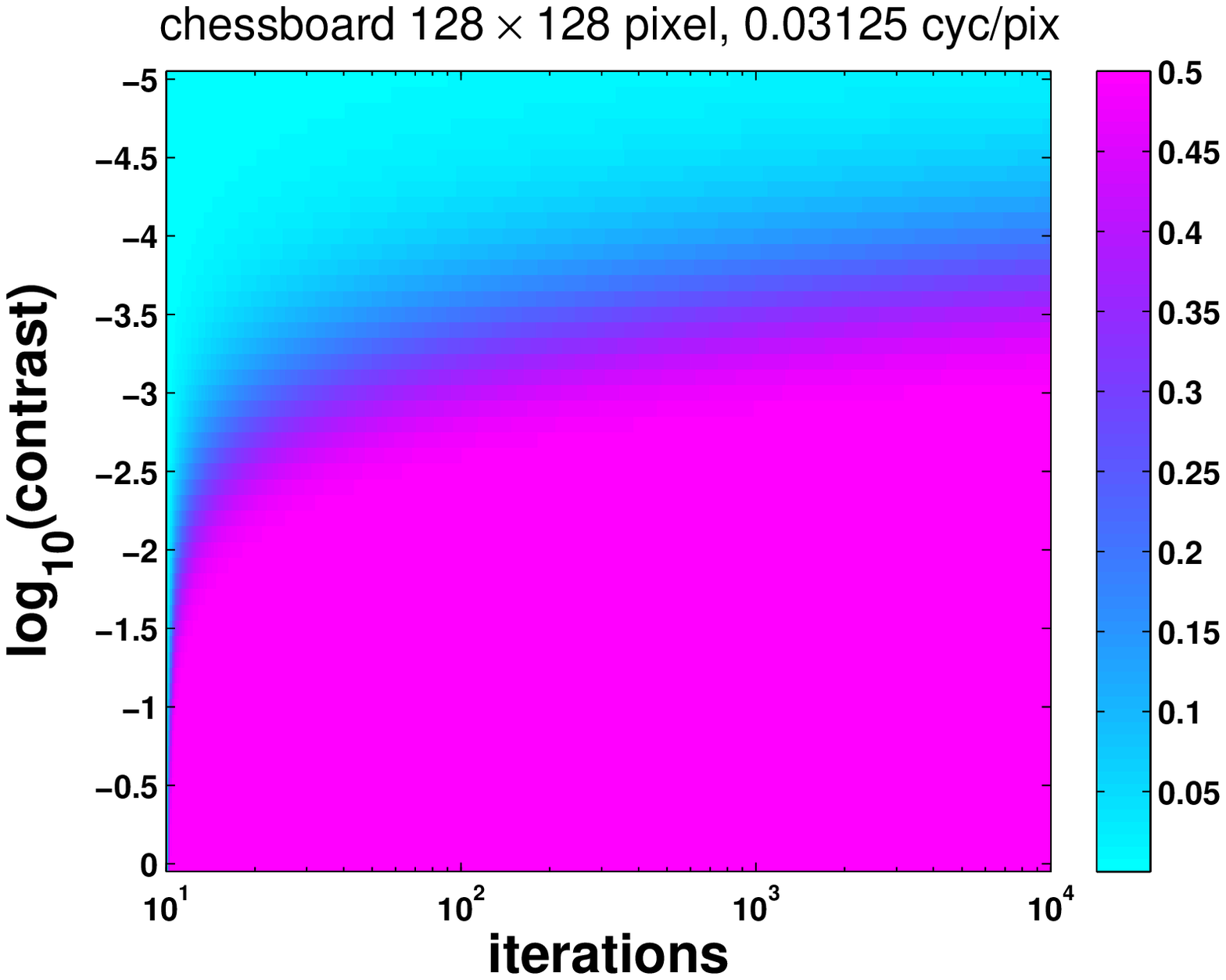}}
	\Caption[minmaxChessContrast][Time to convergence][{%
	Both of the graphs show simulation results of the \dne process with a
	chessboard image as input \size[128].  For both graphs, ``convergence''
	was defined as soon as the average activity of all cells in the \dne
	layer reached $d=0.25$ (note that at full normalization $d=0.5$).  
	{\bf (a)} Time to convergence of the \dne process depends in the first
	place on luminance contrast, and to a lesser extent on spatial frequency content
	of the input (legend: chessboard spatial frequency $k$ in cycles per image)
	-- curves for different spatial frequencies are similar.  The simulation
	results are therefore consistent with \eq[minmaxTau].  Chessboard contrast
	was set to the values indicated on the abscissa: $s_{ij} \in [0,B]$ with
	$B \in \{10^{-5},10^{-4},10^{-3},10^{-2},10^{-1},10^{0}\}$.
	{\bf (b)} Mean activity of the \dne layers is indicated by colors
	(inset: colorbar) as a function of iterations (abscissa) and luminance
	contrast (ordinate) of the chessboard image ($4$ cycles per image).
	The dynamic of the normalization process reveals a sigmoidal behavior which
	consists of a plateau with low activity (top, turquoise), a relatively short
	rising phase (blue), and a plateau with high activity (pink, bottom), where
	convergence has occurred.}]
\end{figure}
%
\subsection{\label{Time2Convergence}Time to convergence for \dne\bs}
%
\Fig[minmaxChessContrast] shows the relationship between the time to
convergence and the numerical range of the input $s_{ij}$: the smaller the
$s_{ij}$, the more iterations are necessary to accomplish the mapping
expressed by \eq[minmaxMapping].  Mathematically, this can be seen as
follows.  Assume that a general solution of \eq[minmaxDynamicNorm] has
the form
\formula[minmaxGeneralSolution][{c_{ij}(t)=C_0\ e^{-t/\tau} + C_1}]
where $C_0$ and $C_1$ are constants which are defined by the initial
conditions, and $\tau$ is a time constant.  Plugging the last equation
into \eq[minmaxDynamicNorm] yields
\formula[minmaxGeneralReplace][{C_0 e^{-t/\tau}
				\left[
					b_{ij}-a_{ij}-\frac{1}{\tau}
				\right]
				= -C_1 \left( b_{ij}-a_{ij} \right)
				+ s_{ij} - a_{ij}}]
By identifying
\formula[minmaxTau][{\tau =\tau_{ij}(t) \equiv  \frac{1}{b_{ij}(t)-a_{ij}(t)}}]
we obtain
\formula[minmaxConstants][{C_1 = \frac{s_{ij}-a_{ij}}{b_{ij}-a_{ij}},}]
and from the initial condition $c_{ij}(t=0)=0\ \ \forall i,j$ we furthermore
get $C_0 = -C_1$, which finally gives the solution
\formula[minmaxSolution][{c_{ij}(t)= \frac{s_{ij}-a_{ij}}{b_{ij}-a_{ij}} 
			\left( 1 - e^{-t/\tau_{ij}} \right).}]
On grounds of the definition of $\tau$ (\eq[minmaxTau]) we obtain two
insights.\\
First, since the time constant $\tau$ of the \dne process is a function
of both $a_{ij}(t)$ and $b_{ij}(t)$, it is not really a constant, but rather
depends on time and space because of \eq[minmaxDiffMin] and
\ref{minmaxDiffMax}, respectively.  However, $\tau$ can be approximated by
recalling that $a_{ij}(t)$ and $b_{ij}(t)$ converge in time and space to
the global minimum $A$ and global maximum $B$, respectively, of the input
$s_{ij}$ (\eq[MINmax] and \ref{minMAX}).  Thus, $\tau \approx 1/(B-A)$.
This leads to the second insight: the smaller are $A$ and $B$, the
longer it takes \dne to converge to a steady state.  Or, otherwise
expressed, the smaller $\tau$ is, the faster the system converges.\\
Notice that when using the steady-state solution (\eq[minmaxLinearScaling])
of \dne instead of the full dynamic process (\eq[minmaxDynamicNorm]), no
dependency on input contrast is revealed, and the dependence on spatial
frequency structure of the input is much weaker.
\begin{figure*}
	\scalebox{0.75}{\includegraphics{\pics/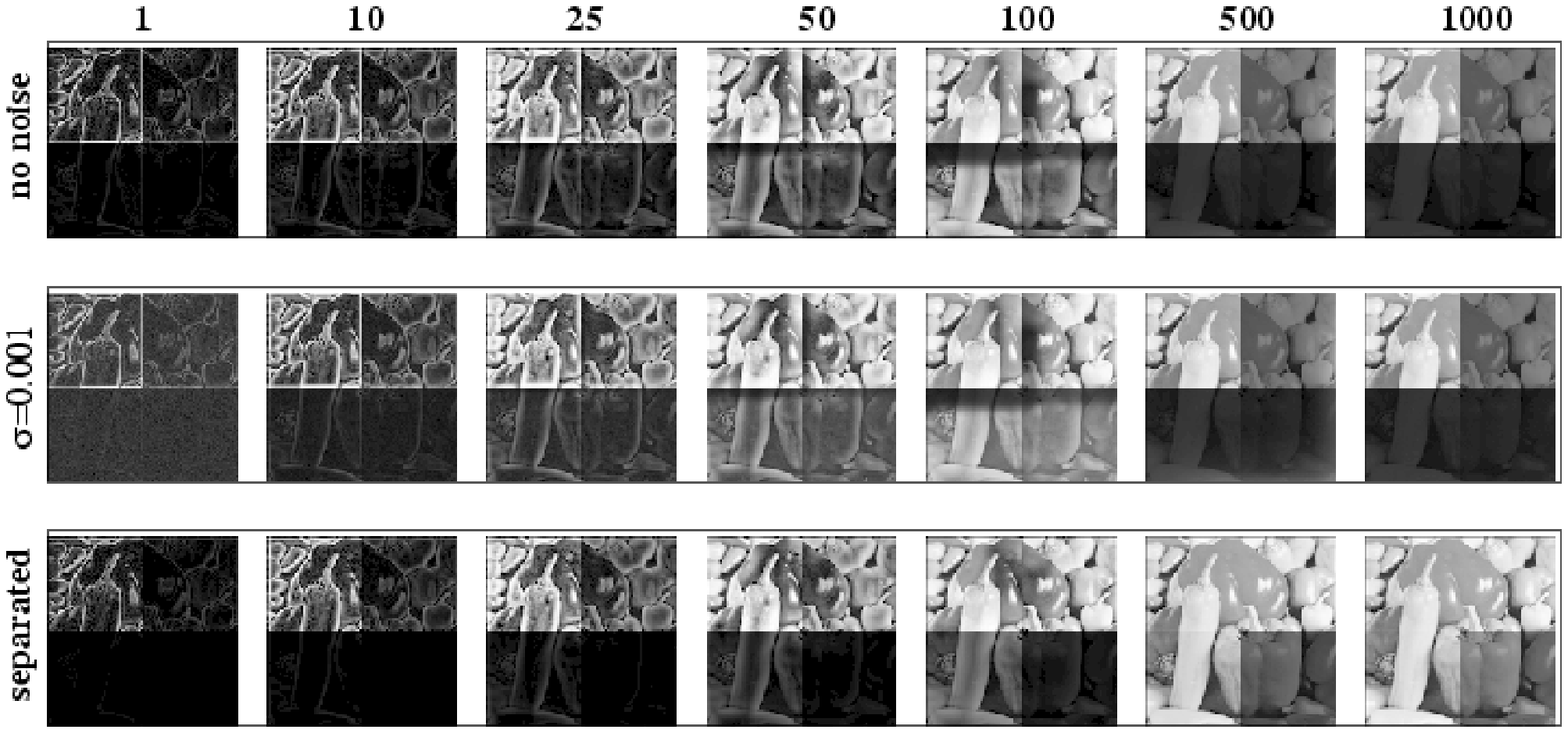}}
	\Caption[dnePeppersTiles][Dynamic compression][{Same as
	\fig[dneLenaPeppers], but here for a ``contrast-tiled'' version of
	\Peppers\bs.  Contrast-tiling means that the original image (entropy 8
	bits, tiled image $\gtrsim 14$ bits, see
	\fig[DynamicCompressionEntropy](a)) was subdivided in quadrants (``tiles'')
	to obtain the dynamic range of luminance values found in a typical outdoor
	scene (values were taken from \cite{JobsonRahmanWoodell97}, table 1).
	At an intermediate number of iterations (around 100), dynamic range
	compression occurs, where details in the darker tiles 
	get better visible.  {\bf Top row}: Noise-free dynamic - compare with
	\fig[DynamicCompression].  The input image is shown as
	inset in \fig[DynamicCompression](b).  {\bf Middle row}:  Dynamic for
	additive Gaussian noise (see \sec[AdditiveGaussianNoise]) with standard
	deviation $\sigma=0.001$ and zero mean - see also \fig[DynamicCompressionNoise]).
	Obviously, moderate levels of noise improve dynamic compression and
	thus the visibility of the darker tiles.
	{\bf Bottom row:} The tiles were disconnected from each other (i.e.
	no diffusion could take place across different contrast tiles), and the
	\dne mechanism now treats each tile as a separate image.   Notice that
	in the first two rows all four tiles are connected (i.e., the tiles are
	treated as a single image), and activity propagates between tiles (as it
	is visible by a black shadow in the lower tiles at around $100$ iterations).}]
\end{figure*}
\begin{figure}
	\scalebox{\figscale}{%
			\includegraphics{\pics/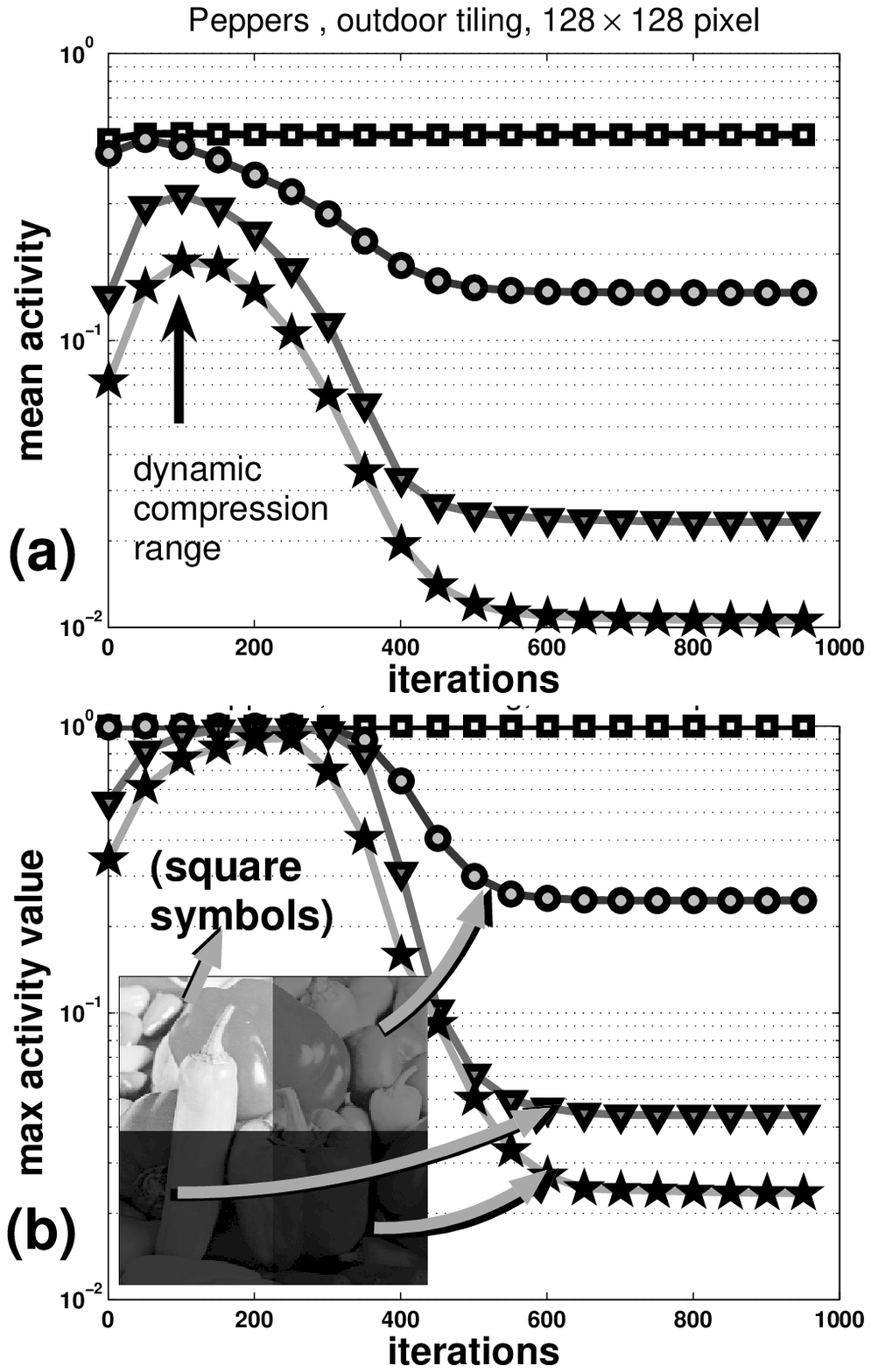}
			}
	\Caption[DynamicCompression][Dynamic compression of input range][{%
	Each curve quantifies the activity across one of the four tiles (inset)
	of the dynamic shown in the top row of \fig[dnePeppersTiles]:
	{\bf (a)} mean activity, and  {\bf (b)} maximum of activity.
	Within the time window where dynamic compression is seen, initially
	separated curves approach each other, and subsequently depart again.
	Other real-world images give similar results.}]
\end{figure}
\begin{figure*}
	\scalebox{0.75}{\includegraphics{\pics/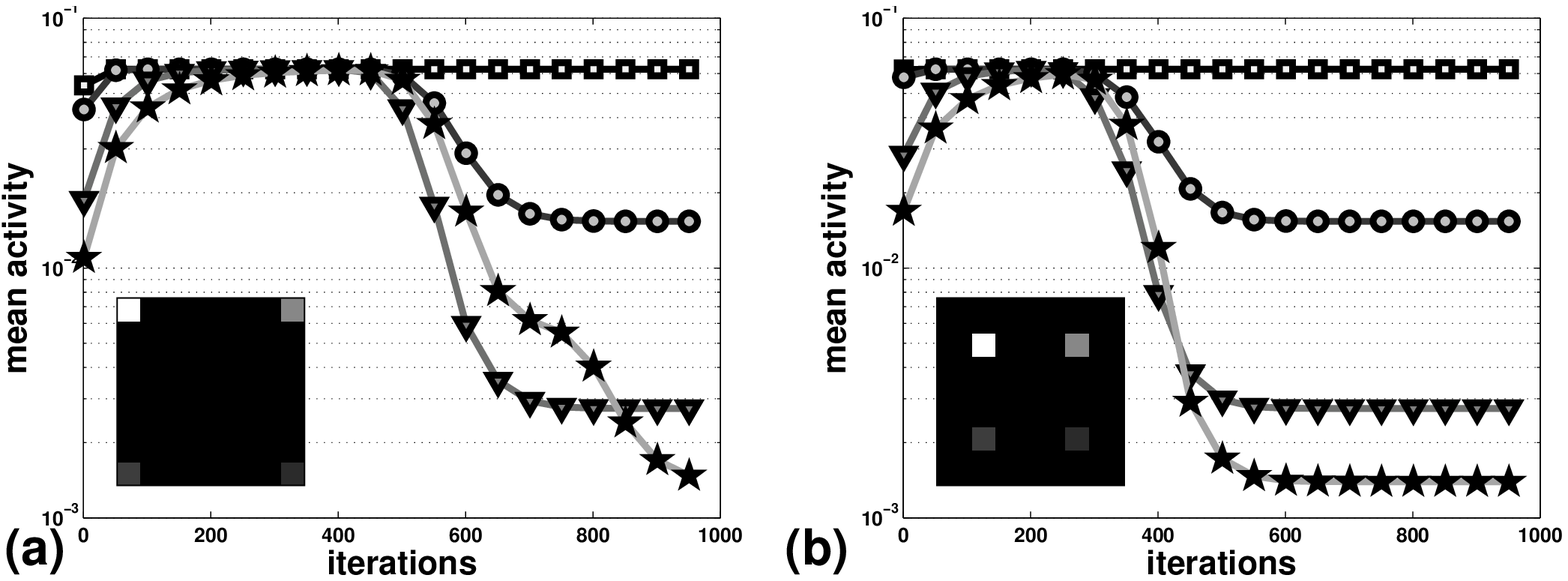}}
	\Caption[Understanding][Understanding dynamic range compression][{%
	The dynamic compression effect is a consequence of that the global
	maximum propagates with finite speed in the \masyx.
	Thus, the smaller the initial region occupied by the global maximum
	in the \masyx, and the greater the distance of this region from other
	regions of smaller cell activities or local maxima, respectively,
	the longer the persistence of the effect.  This is illustrated with
	two images (insets).  The images consist of zeros except for small
	squares with different luminance values.  The global maximum corresponds
	to the white square ($s=1$, upper left in the images).  The luminance
	values of the squares are the same as the maximum value of each quadrant
	in the tiled \Peppers image of \fig[DynamicCompression], that is $s=1$,
	$s=0.24615$, $s=0.04385$, and $s=0.02231$, respectively.
	{\bf (a)} The squares are maximally separated, and the global maximum
	reaches the other local maxima relatively late (mean activities were computed
	across the same tiles as before with the tiled \Peppers image).  As a
	consequence, each square is independently normalized by its local
	maximum until it gets invaded by a higher activity value. 
	{\bf (b)}  Moving the four squares closer to each other goes
	along with a shorter duration of the dynamic compression effect.
	Were all four squares moved into the center of the image such
	that they touch each other, virtually no dynamic compression
	effect would be revealed, because the global maximum would
	instantaneously propagate to all four tiles: all local maxima would
	get normalized by the global maximum right off.}]
\end{figure*}
%
\subsection{\label{section_DynamicCompression}%
			Transient adaptation or dynamic compression}
The \dne layer reveals distinct dynamic phases.
In the initial phase, image contrasts are extracted.  Contrast enhancement
occurs in a subsequent phase.  In the final phase, the activity distribution
in the \dne layer is just a re-scaled version of the input.  In a second phase
between the initial and the final phase, one observes adaptation: image
structures with substantially different light intensities in the input are
mapped to a smaller range of activities in the \dne layer.   This effect is the
\emph{dynamic range compression}.  For its illustration an input image
was subdivided into four quadrants (``contrast tiles'', \fig[dnePeppersTiles]).
Each of the tiles has a different range of luminance values.  Because the 
available tonal range for displaying the tiled image is too small to
match the range of all tiles, some of the image details in the darker
tiles are displayed in black.  Nevertheless, a part of these details
are rendered visible in the \dne layer at around 100 iterations
(top row in \fig[dnePeppersTiles]), implying that cell activities
in this layer have less dynamic range than in the input.  The compression 
effect is quantized in \fig[DynamicCompression], where each curve represent
the mean activity and the maximum activity, respectively, of all cells of one
the four tiles.  The curves approach each
other at around 100 iterations.  Thus, the output of the \dne network can be
encoded with a smaller than the original numerical range.\\
\Fig[Understanding] illustrates the mechanism which underlies dynamic compression.
A necessary condition for dynamic compression to occur is that the global maximum
propagates with finite speed in the \masyx, and that it is spatially separated
from image structures that have less dynamic range ($=$ local maxima).  When
the global maxima has not yet propagated to the local maxima, then image
structures are normalized by their ``own'' local maxima.  Since normalization
rescales all cell activities to the same target range (all image structures
normalize to one), local normalization implies a reduction of the dynamic range.
However, local maxima are annihilated as the global maximum propagates, and image
structures are now getting normalized by the global maximum.  Then, the entire
dynamic range of the input image is recovered in the normalization layer, and
dynamic range compression is abolished.  The recovery of the original dynamic
range can be seen when the entropy curves of \fig[DynamicCompressionEntropy](b)
reduce to the entropy of the input image at $\approx 1000$ iterations (dashed
horizontal line).
\begin{figure}
	  {\bf \large (a)}\scalebox{0.45}{\includegraphics{\pics/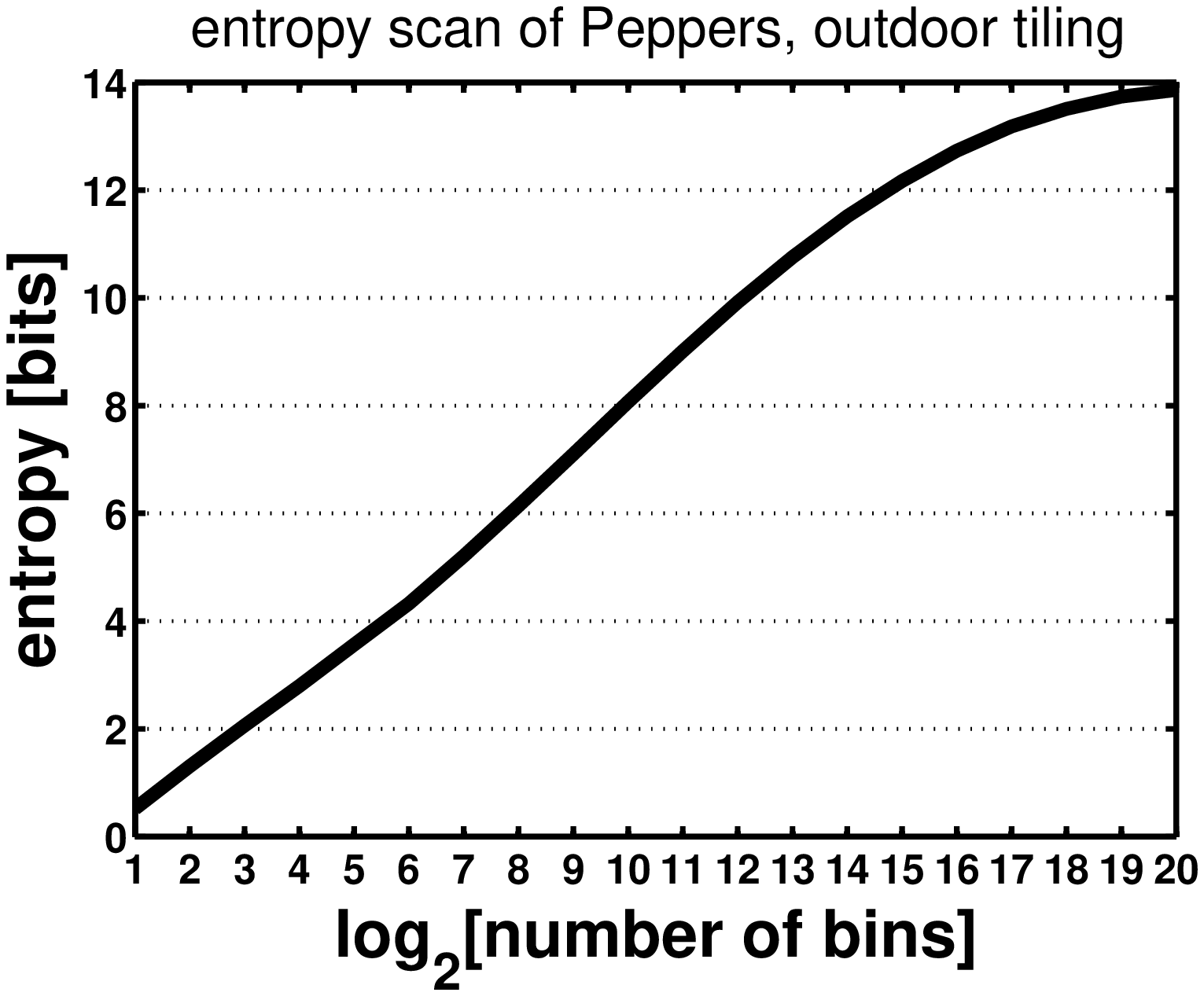}}
	  {\bf \large (b)}\scalebox{0.45}{\includegraphics{\pics/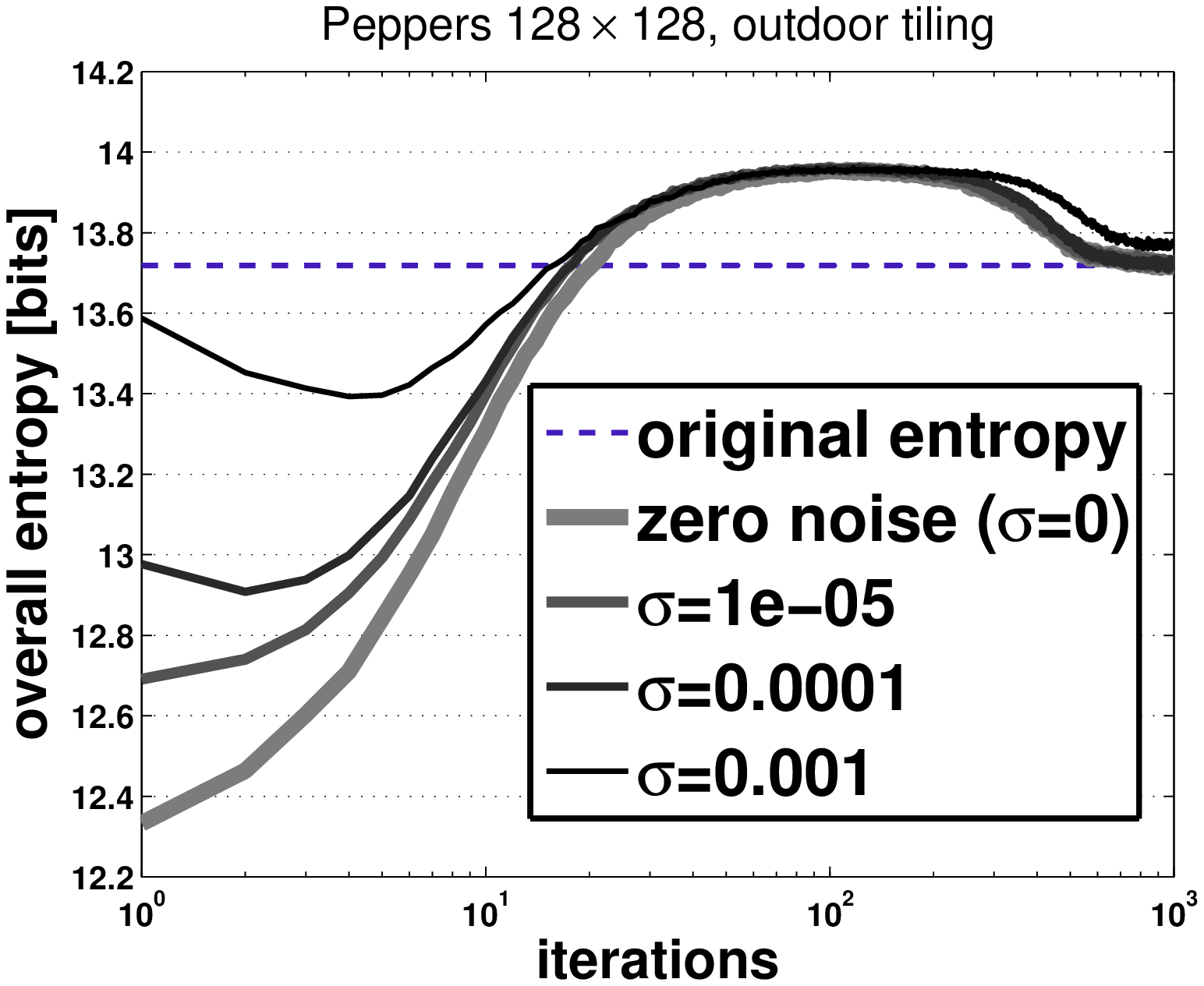}}
	\Caption[DynamicCompressionEntropy][Dynamic compression, entropy and Gaussian noise][{%
	{\bf (a)} ``Entropy scan'' of the tiled \Peppers image.  Shannon
	entropy \cite{Shannon48} was computed as a function of the number
	of histogram bins.  The curve starts to saturate at approximatively
	$10^{20} = 1048576$ bins (the maximum value allowed with the computer
	that was used for the simulations).  Thus, the entropy of the tiled
	image is $\gtrsim 14$ bits.
	{\bf (b)}  Shannon entropy of the \dne layer a function of iterations.
	The tiled \Peppers image served as input.  Here, hardware constraints
	only permitted the computation of entropy with $15\cdot 10^5$ bins.
	Each curve represents a different amount of temporally varying
	and normal-distributed noise (additive ``Gaussian noise'') with 
	standard deviations $\sigma \in \{0,10^{-5},10^{-4},10^{-3}\}$
	(see inset).   The dynamic compression effect is associated with
	plateau-like maxima in the entropy curves.  The dashed horizontal
	line denotes the entropy of the input image.}]
\end{figure}
\begin{figure*}
	\scalebox{0.75}{%
	    \includegraphics{\pics/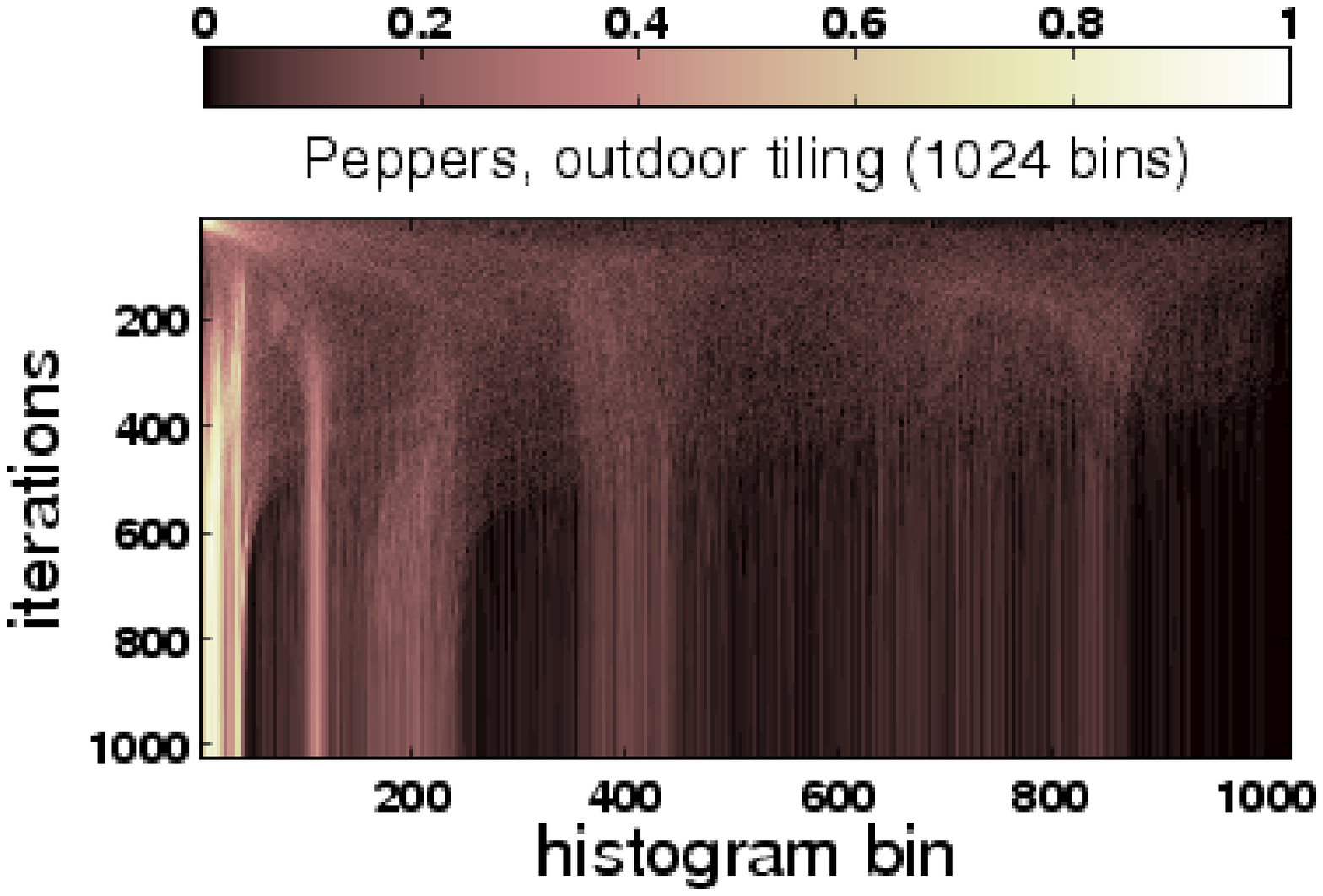}
	    }
	\Caption[DynamicCompressionHistogram][Histogram evolution
	of the tiled \Peppers image][{Each horizontal line in the
	above image represents a histogram of the \dne layer (abscissa:
	$1024$ bins per line) at a different iteration number (ordinate).
	Occurrence frequencies of output activity values are represented
	by colors (inset: colorbar).\\
	Initially, cells in the \dne layer have small activities
	and concentrate in the first histogram bins (upper left corner;
	the first  14 time steps were dropped for visualization reasons).
	This highly predictable state is associated with low entropy.
	Immediately afterwards, values distribute themselves homogeneously
	over virtually all bins, and thus entropy increases.  This is when
	dynamic compression occurs:  an observer who is monitoring the output
	of the \dne network gathers the highest information about the input
	image.  In other words, dynamic range compression is compatible with adaptation,
	since adaptation maximizes the transfer of information \cite{Wainwright99}.
	Subsequently values are redistributed again in a way that they concentrate
	along four principal stripes reflecting the four contrast tiles.  This
	state is again associated with a lower entropy.
	Notice that both the way from and to the more homogeneous
	distribution of values is mirrored in structures similar to faint
	``trajectories'' that sweep from left to right across the histogram.}]
\end{figure*}
%
%
\subsection{\label{section_ProcessEntropy}%
			Process entropy}
%
%
%
\Fig[DynamicCompressionEntropy](b) shows entropy as a function of time computed
over the \dne layer.  The entropy reaches a maximum in the time window where
dynamic compression occurs.  Notice that this maximum in entropy exceeds the
entropy of the input image (dashed horizontal line).  Because entropy
quantifies the degree of flatness of a histogram (or probability distribution),
the observed entropy maximum implies that cell activities of the \dne layer are
more homogeneously distributed across the histogram than luminance values of
the input. \Fig[DynamicCompressionHistogram] shows how the distribution of
activities evolve over time.  Initially, cell activities in the \dne layer
are small, and tend to cluster around a single spot in the histogram
(the cropped ''hot spot'' in the upper left corner of the histogram).
Emanating from this ``hot spot'', values start to occupy nearly the
entire histogram.  It is just then when an observer who is monitoring
the output of the \dne network gathers the highest information about
the input image.\\
In the consecutive part of the dynamic, the values are redistributed again
in a way that they concentrate around four principal stripes.  These stripes
correspond to the four contrast tiles.  Therefore, dynamic range compression
is compatible with adaptation, since adaptation maximizes the transfer of
information \cite{Wainwright99}.
\begin{figure}
	  \scalebox{0.45}{\includegraphics{\pics/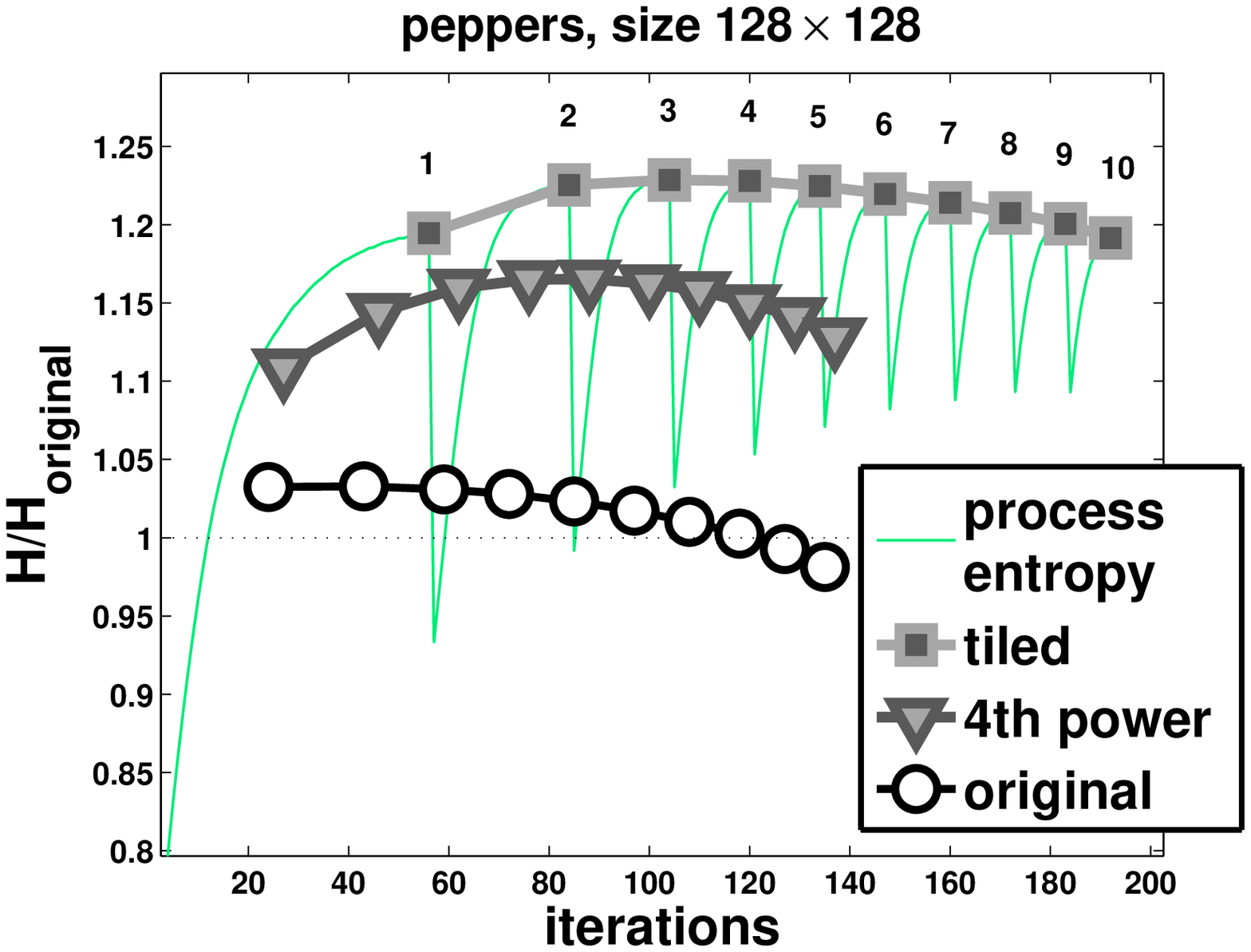}}
	\Caption[AdaptationEntropy][Adaptation by entropy maximisation (entropy)][{%
	Each point of the curves ``\emph{original}'', ``\emph{4th power}'', and
	``\emph{tiled}'' corresponds to the entropy $H$ (normalized by the
	entropy of each respective original image $H_\mathrm{original}$) at
	the indicated number of feedback loops of the entropy maximizing
	adaptation algorithm (output images at one, two, three and 20 loops
	are shown in the previous figure).  In addition, the curve designated
	by ``\emph{process entropy}'' relates the relative entropy of the full
	algorithm to the data points of the ``\emph{tiled}'' \Peppers image:
	between any two data points, the \dne network was run by using the
	output image at the previous data point as input until a maximum
	in entropy has been reached.  See text for further details.}]
\end{figure}
\begin{figure}
	  \scalebox{0.6}{\includegraphics{\pics/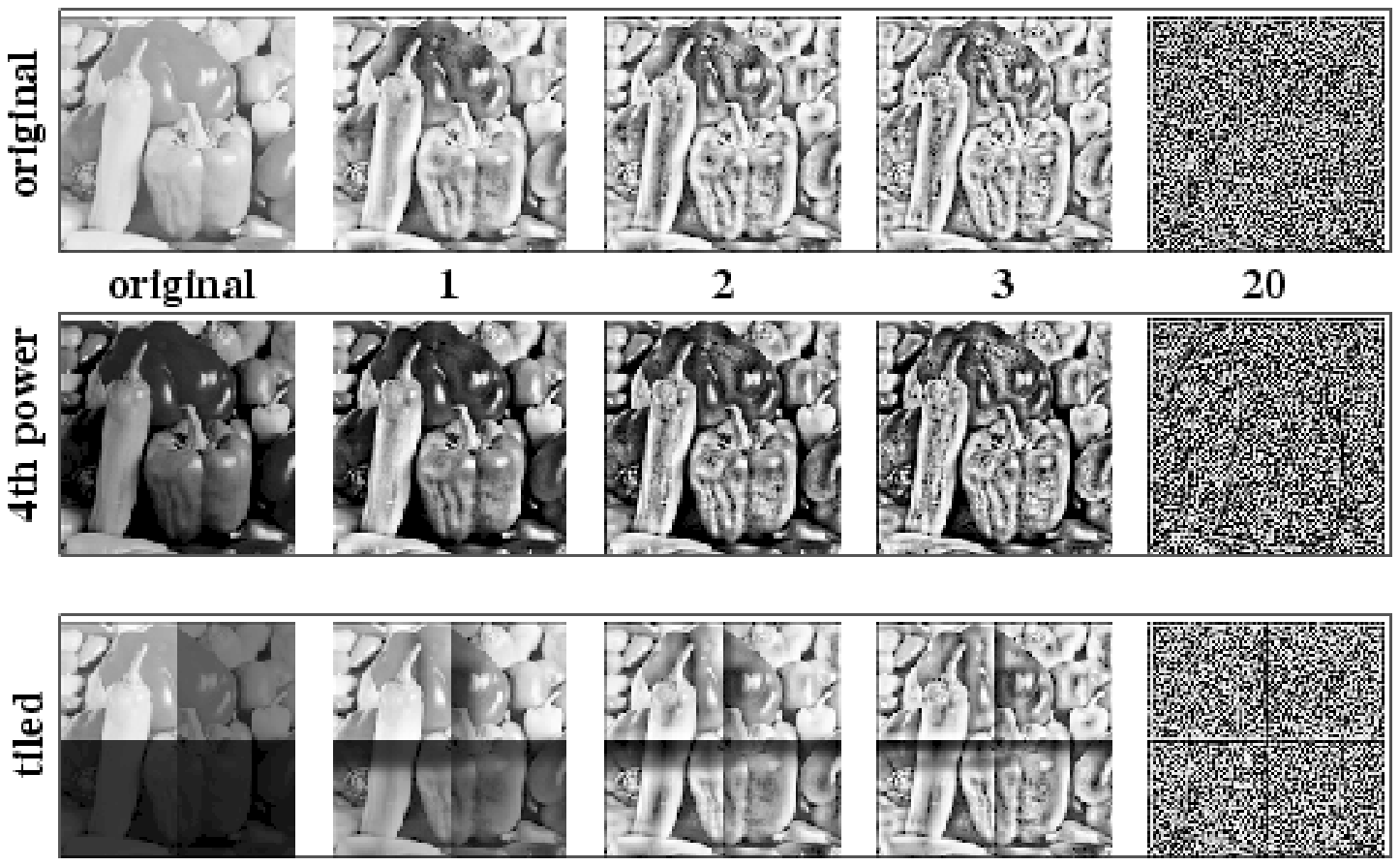}}
	\Caption[AdaptationResults][Adaptation by entropy maximisation (results)][{%
	The first column shows the original images which were used as input to the
	\aem algorithm: ``\emph{original}'' is the original pepper image;
	``\emph{4th power}'' is the original pepper images with luminance values
	elevated by forth power (in this way a high-dynamic-range image is created);
	\emph{tiled} is the tiled peppers image (cf. \fig[dnePeppersTiles]).  The
	numbers designate how many feedback loops of the algorithm were run.}]
\end{figure}
%
%
%
%
%
%
\subsection{\label{section_Adaptation}\Aemx}
%
%
When computing the Shannon entropy of the output of the \dne network
\cite{Shannon48}, one observes an entropy maximum at the dynamic
compression effect (\fig[DynamicCompressionEntropy] and
\ref{DynamicCompressionHistogram}).  Hence, a
straightforward algorithm for the adaptation of images is to stop
the \dne process when an entropy maximum is reached (``one feedback
loop'').  To further enhance the dynamic compression effect, the output
at the entropy maximum can be taken again as input to the \dne network,
and once again we can let the \dne process continue until it reaches a maximum
of entropy (``two feedback loops'').  The entropy across $10$ feedback loops
of the just described algorithm is illustrated in \fig[AdaptationEntropy] with
the curve designated by ``process entropy''.  \Fig[AdaptationResults] shows the
output images obtained for one, two, three and $20$ feedback loops.  With
increasing number of feedback loops, luminance information is suppressed, while
contrasts are enhanced.  At around $20$ loops, one obtains an image which seems
to contain only contours, but iterating further enhances also noise and leaves
one with an image without any recognizable structures.
\Fig[AdaptationEntropy] shows that entropy decreases with increasing number
of feedback loops (each data point is the entropy of the output at the indicated
number of feedback loops).
For the ``\emph{tiled}'' and the ``\emph{4th power}'' \Peppers image,
the entropy versus feedback loops has a maximum.  Concluding, in terms
of entropy, but also by visual inspection, a small number of feedback loops
(one or two) seems optimal for the proposed adaptation algorithm.  The
algorithm should be understood as  a ``proof-of-concept'' rather than
a definite tool for image processing, because it occasionally develops
artifacts.  For example, future versions could address the suppression
of the dark zone which emanates from the tiles of the ``\emph{tiled}''
\Peppers image.

\begin{figure}
	\scalebox{\figscale}{%
		\includegraphics{\pics/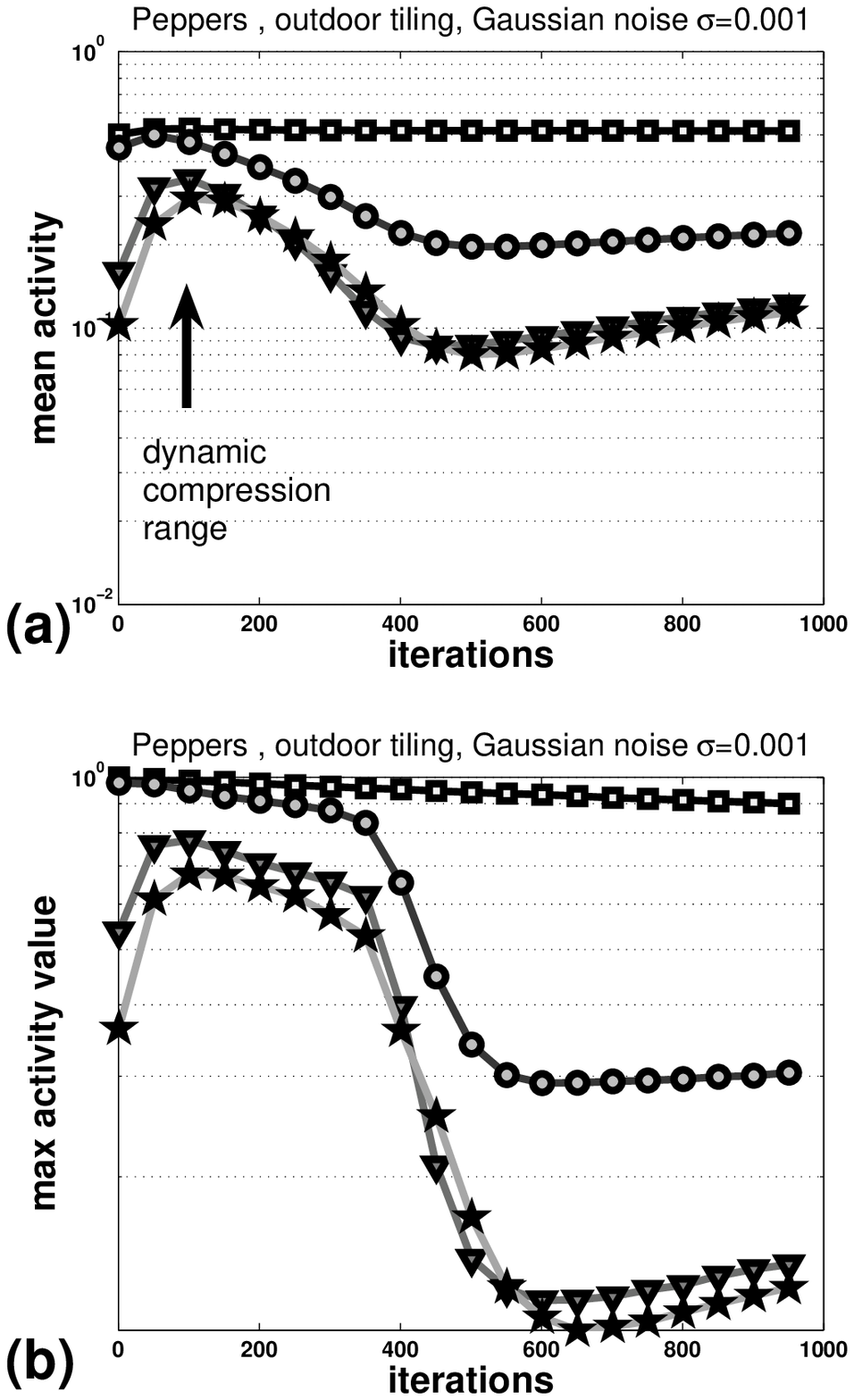}%
		}
	\Caption[DynamicCompressionNoise][Dynamic compression in the presence
	of noise][{Same as \fig[DynamicCompression], but here with additive and
	temporally varying Gaussian noise (standard deviation $\sigma=0.001$,
	zero mean).  Snapshots of the noisy dynamic are shown in the middle
	row of \fig[dnePeppersTiles].  Evidently, suitable chosen noise levels
	can enhance the dynamic compression effect.  See also
	\fig[DynamicCompressionEntropy](b) for the dependence of entropy on
	noise levels.}]
\end{figure}
\begin{figure}
	  {\bf \large (a)}\scalebox{0.45}{\includegraphics{\pics/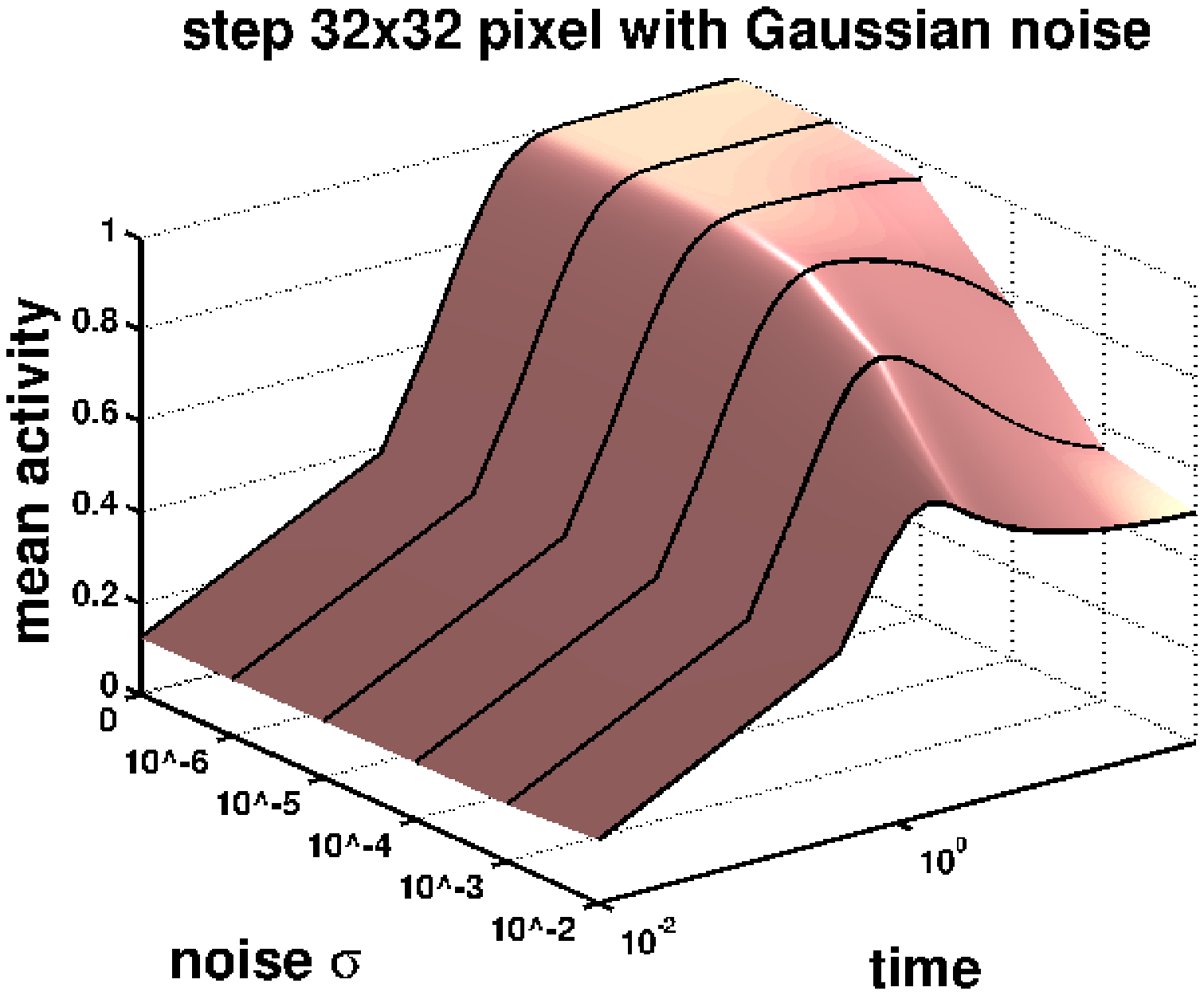}}
	  {\bf \large (b)}\scalebox{0.45}{\includegraphics{\pics/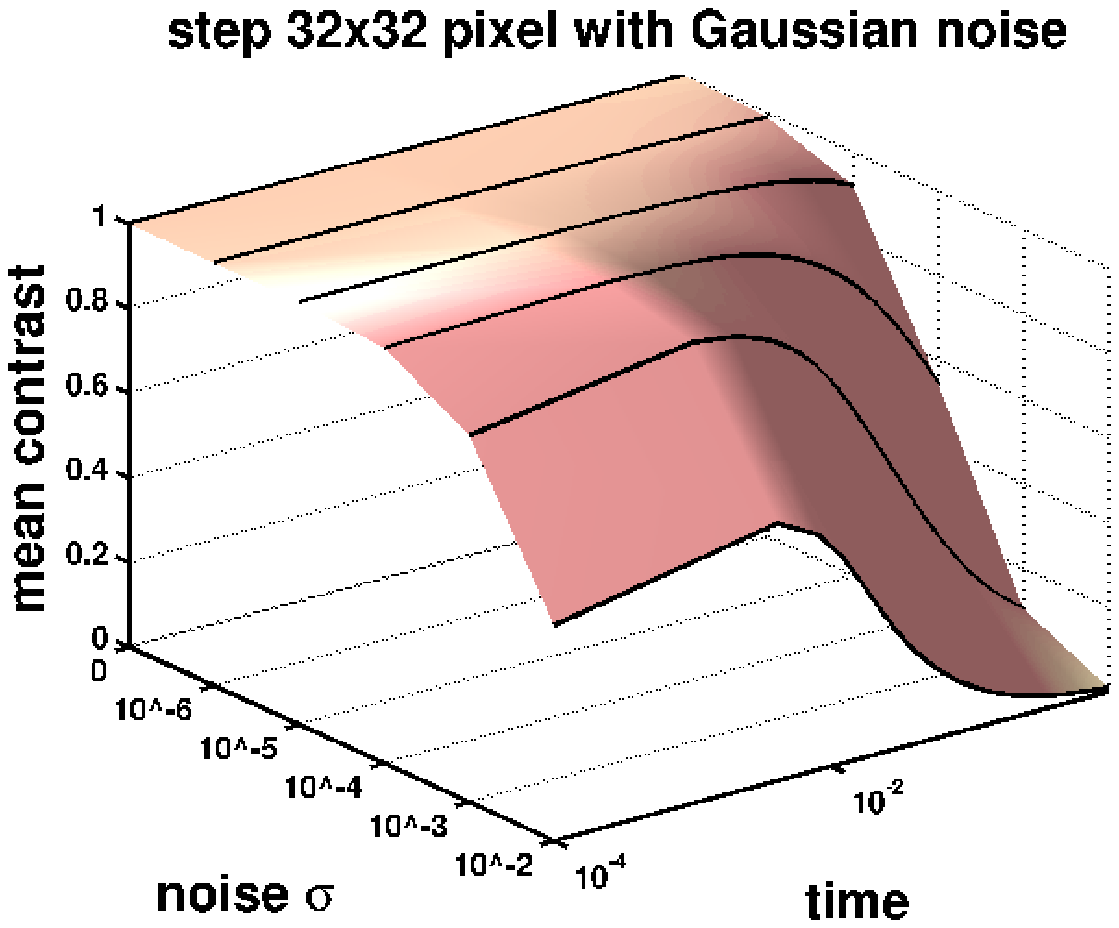}}
	\Caption[Noise][\Dne with additive Gaussian noise][{A luminance step (activities
	zero and one) was used to analyze the behavior of \dne network in the
	presence of additive and temporally varying Gaussian noise (zero mean,
	standard deviation $\sigma$ as indicated in the plot).
	{\bf (a)} Activity of \dne averaged over the cells corresponding to the 
	white region of the luminance step.
	{\bf (b)} Michelson contrast between the black and the white region
	of the step, averaged over respective positions.  See text for further
	details.}]
\end{figure}
%
%
%
\subsection{\label{DynamicNoise}Sensitivity of \dne for noise}
%
One may argue that an adaptive mechanism designed in a way suggested by \dne
is highly sensitive to noise, because it is based on the computation of
minimum and maximum operations.  To address this issue, we further distinguish
between static noise (i.e. an offset added to the input $s_{ij}$ which does not
vary with time), and dynamic noise (i.e. an offset added to each layer which
varies with time).   For the first case we presume the existence of a noise-free
input pattern, to which static noise is added.  A worst case scenario is on hand
if a couple of cells $s_{ij}$ have high activities due to noise (``noisy cells'')
which lead to an undesired increase of the true dynamic range of the input.
If a read-out mechanism for the \dne layer had only the same dynamic range
as the input, then the noise would obscure the relevant information of the
input at convergence.  Nevertheless, if there were only a few noisy cells in
the input layer, then the dynamic compression effect could mitigate the
worst case scenario to some extent.\\
To assess the robustness of \dne against temporally varying noise, numerical
experiments were conducted with additive, normal-distributed noise (``Gaussian
noise''), with zero mean and standard deviation $\sigma$.  Apart, additional
simulations were conducted with multiplicative, uniform noise (``white noise'').
\subsubsection{\label{AdditiveGaussianNoise}%
			Additive, normal-distributed noise}
Temporally fluctuating normal-distributed noise $\xi_{ij}(t)$ was added to
the equations \ref{minmaxDiffMin}, \ref{minmaxDiffMax}, \ref{minmaxDynamicNorm},
and the input $s_{ij}$, according to
\formula[GaussianNoise][{x_{ij} \leftarrow x_{ij} + \sigma \xi_{ij}(t).}]
\def\avrg[#1][#2]{$<\!\!\!#1\!\!\!>_{#2}$}
In the last equation, $x_{ij}$ stands for one of the variables $a_{ij}$,
$b_{ij}$, $c_{ij}$, and $s_{ij}$, respectively, and ``$\leftarrow$''
means that the left hand side is replaced by the right hand side.  The
noise level is specified by $\sigma$ (assuming zero mean), and $\xi_{ij}(t)$
is assumed to be not correlated across time and/or spatial positions.  A
luminance step (32 $\times$ 32 pixels) was used as input, with luminance
value zero on the dark side (``black patch'', columns 1 to 16), and 1 on the
bright side (``white patch'', columns 17 to 32).  Thus, the mean activity of
the noise free system should approach one at steady-state.  We furthermore
computed the Michelson contrast $\mathcal{M}$ at each position $(i,j)$
according to
\formula[Michelson][	\mathcal{M}_{ij} = {\frac{c_{ij\in\mathrm{white}}
			-c_{ij\in\mathrm{black}}}
			{c_{ij\in\mathrm{white}}+c_{ij\in\mathrm{black}}}}]
where $j\in\mathrm{black}$ means $1 \leq j \leq 16$ and $j\in\mathrm{white}$
means $17 \leq j \leq 32$ (the row index runs over all positions
$1 \leq i \leq 32$).\\
\Fig[Noise](a) shows the temporal evolution of the mean activity of
\avrg[c_{ij}][ij\in\mathrm{white}] (i.e. averaged over white patch
positions) for various noise levels $\sigma$.  Sufficiently high noise
levels significantly affect the convergence behavior of \dne - the
response plateau which is seen in the noise-free case is no longer
reached.  Instead of the plateau, a maximum is approached, the amplitude
of which decreases with increasing noise level.  \Fig[Noise](b) shows
that a similar behavior is also seen for the averaged Michelson-contrast
\avrg[\mathcal{M}_{ij}][]: The contrast between the black and the white
patch decreases with increasing noise level.  This implies that image
structures are obscured by noise.\\
How does noise take influence on dynamic range compression?  Three answers
exist to this question, and they depend on the noise level.  For relatively
small noise levels ($\sigma < 0.001$), no dramatic effect on dynamic compression
is observed.  For intermediate noise levels ($\sigma \approx 0.001$), dynamic
compression is enhanced (bottom row in \fig[dnePeppersTiles], and
\fig[DynamicCompressionNoise]).  Enhancement happens because the net
effect of noise is to add an offset, which ``lifts'' the darker patches of
the tiled \Peppers image.  For higher noise levels, however, the darker
patches drown in noise and image details get lost.  Consequently, if the
goal of \dne was adaptation, then suitable chosen noise levels would aid to
enhance range compression, although this comes at the prize of reduced
contrasts in regions with low activities (dark quadrants in the tiled
\Peppers image).\\
Notice that additive Gaussian noise can be easily counteracted by proposing
additional mechanisms with low-pass characteristics, like spatial or temporal
pooling of activity.  Then, as long as the noise is not correlated over positions
and time, it would simply average out.
\subsubsection{Multiplicative and normally distributed noise} 
Multiplicative noise was applied to variables $a_{ij}$, $b_{ij}$, $c_{ij}$,
and $s_{ij}$, respectively, according to
\formula[MultNoise][{	x_{ij} \leftarrow x_{ij} \cdot
			(1 + \eta *(\mu_{ij}(t)-1)) }]
with $0 \leq \mu_{ij}(t) \leq 1$ representing uniformly distributed noise
which was uncorrelated across time and/or space.  The noise level is specified
by $\eta\in[0,1]$.  \Dne is not significantly affected by this type of noise,
not even for $\eta=1$ (hence results are not shown).  Multiplicative noise
acts differently on maxima and minima.  Maximum activities can only decrease,
but never increase beyond their value in the noise free case.  Therefore, no
spurious maxima are introduced into the \masy by the type of multiplicative
noise considered here.  On the other hand, multiplicative noise can inject
spurious minima into the \misyx, if the lowest luminance value in the input
image was bigger than zero.  As the minimum luminance values of our images
were always zero, they are consequently not affected by the multiplicative noise. 
%
%
%
\section{Discussion}
%
%
%
\subsection{\Pseudox-diffusion and electrical synapses (gap junctions)}
%
%
The operator $\KEILIAN[\lambda]$ models different types of electrical synapses
(gap junctions).  In its linear version, $\KEILIAN[\lambda=0]$ describes the
exchange of both depolarizing (i.e. directed towards a neuron's firing
threshold) and hyperpolarizing (i.e. directed away from a neuron's firing
threshold) currents between adjacent neurons.  Networks of electrically
coupled neurons are ubiquitous both in the retina (e.g.
\cite{RaviolaGilula75,Kolb77,NelsonEtAl85,MillsMassey94,MillsMassey00})
and the cortex (e.g. \cite{GiBeCo99,GalHes99,GalHes02}).
These networks can be modeled by diffusion equations (e.g.
\cite{NakaRushton67,Lamb76,Winfree95,BendaEtAl01}).  Conversely, the
operators defined by \eq[minmaxCompactPos] and \ref{minmaxCompactNeg}
represent models for rectifying (i.e. voltage sensitive) gap-junctions.
Rectifying gap junctions were described in the crayfish (e.g.
\cite{Edwards91,Edwards98}), and unidirectional and gated gap junctions
were reported in the rat (e.g. \cite{BukauskasEtAl02}) and turtle
(e.g. \cite{PiccolinoEtAl84,BukauskasEtAl02}), respectively.\\
In organisms, rectifying gap junctions may nevertheless be
implemented in a ``dirty'' fashion.  This means that a current flux
may not strictly occur in only one direction.  Rather, a small amount
of current may as well flow in the opposite direction.  Such behavior
is captured by setting $\lambda$ to a finite value $1 \ll | \lambda | < \infty$,
and was analyzed in \fig[TwoCellSystem]. 
%
%
%
\subsection{Computational aspects}
%
%
%
Substitution of two global memories (for the minimum and the maximum activity)
by two \pseudox-diffusion layers of size $N\times N$ leads to a computationally
more demanding system, because more memory resources are needed and significantly
more computational operations need to be carried out for their simulation.  Moreover,
because computation of the global maximum or minimum is based on local, diffusion-like
interactions, a maximum or a minimum does not propagate from one cell to another
from one time step to the next.  The diffusion rate cannot be chosen arbitrarily
high to guarantee the numerical stability of the process.  The time to convergence
does not only depend on the \pseudox-diffusion layers reaching a steady-state,
but is mainly determined by the \dne layer.  The number of iterations that is
needed until convergence occurs scales with the numerical range of the input.
Thus, for small input values, the number of required iterations can be quite
large (see \fig[minmaxChessContrast]).  Therefore,
the \dne network cannot be seriously considered as an alternative to an
ordinary normalization algorithm (i.e., searching the global maximum and
minimum, and then rescaling).  However, the \dne network can accomplish
different tasks which cannot be accomplished with an ordinary normalization
algorithm, for example detection of contrast contours, or compression of
the dynamic range of the input.
%
%
\section{Conclusions} 
%
This paper introduced a parameterized diffusion operator (parameter $\lambda$)
and analyzed some of its properties mathematically and by computer simulations.
As a special case, heat diffusion is obtained for $\lambda=0$.
Diffusion layers which are based on the two limit cases of the operator (for
$\lambda\rightarrow\pm\infty$) compute the global maximum and minimum, respectively,
of the initial cell activities of the layer.  This means that at convergence, all cells
of the diffusion layers contain the same activity value -- the maximum
($\lambda\rightarrow\infty$) or the minimum ($\lambda\rightarrow - \infty$).
Based on these operators, a \dne network was defined (\eq[minmaxDynamicNorm]).
Its steady-state solution is functionally equivalent to the ordinary rescaling of
a set of numbers (\eq[minmaxLinearScaling]), but by making the normalization
process dynamic, one observes two additional properties: contrast enhancement
and dynamic range compression.
Both effects occur because at first normalization acts locally, similar to
adaptation mechanisms.  With increasing time, the normalization process gets
continuously more global, until a steady-state is reached.  The steady-state
corresponds to a rescaling of the input in the \dne layer.\\
By exploiting the dynamic compression effect, it should thus be possible to
design a powerful adaptation mechanism which maps an input image of an
arbitrary numerical range to a smaller target range.  To do so, the
normalization process has to be ``frozen'' when dynamic compression occurs.
As a first step into that direction, a simple adaptation algorithm based on
the maximisation of entropy was proposed (\sec[section_Adaptation]): the dynamic
is frozen as soon as a maximum of entropy is reached, and the output is 
then fed back as new input to the \dne network.  As a further improvement,
the diffusion operators could be modified such that activity exchange between
two cells is blocked for sufficiently large activity gradients \cite{MatsEtAl05}.
Doing so would possibly prevent in \fig[dnePeppersTiles] (first and second row) the global
maximum from spreading between tiles, and would normalize each tile independently,
such that ideally a dynamic similar to the bottom row in \fig[dnePeppersTiles]
is produced.\\
Systems based on \pseudox-diffusion have already turned out to be of utility
for a variety of purposes in image processing (for implementing filling-in
mechanisms, or winner-takes all inhibition, see \cite{MatsEtAl05}).
\Pseudox-diffusion systems can generally be used for implementing the max-operation
without the need for globally acting pooling units (see for example
\cite{YuGiesePoggio02} and \cite{MatsThesis}).  The advantage over
functionally equivalent but hardwired systems is that the region where
normalization takes place can be dynamically adjusted.  Furthermore,
the maximum operation serves to implement invariance properties in
models for object recognition (e.g. \cite{RiesenhuberPoggio99,RiesenhuberPoggio00}).
%
%
\begin{acknowledgments}
This work was supported by the \emph{Juan de la Cierva} program of the Spanish
government, and the the MCyT grant SEJ 2006-15095.  Further support was provided
by the AMOVIP INCO-DC 961646 grant from the European Community.  The author
likes to thank the anonymous reviewer of \emph{Physica D} for his valuable
suggestions which helped to improve the manuscript (the present version is
the long version; a shorter version is to be published in \emph{Physica D}).
\end{acknowledgments}
\appendix
\section{\label{Methods}Material and methods}
All simulations were carried out using the Matlab environment (R2006b)
on a Linux workstation, where both native Matlab code and mex-files
programmed in C++ were used.  Diffusion operators were normalized
by the number of adjacent cells (normally four, along the domain
boundaries three, and in the corners two).
Normally, the equations describing the diffusion layers
(eqs. \ref{minmaxDiffMin} and \ref{minmaxDiffMax}), and \dne (eq.
\ref{minmaxDynamicNorm}) turned out to be numerically stable such that
a forward-time-centered-space (FTCS) Euler scheme with step size
one is sufficient. (Here, we understand numerical stability such that
the solution converges rather than growing in an unbounded fashion).
Notice, however, the stability criterion associated with the FTCS-integration
of the heat diffusion equation $D\cdot \Delta t \leq 1/2$ (assuming grid
spacing one, see section 19.2 in \cite{Recipes}) where $D$ is the diffusion
coefficient, and $\Delta t$ is the integration step size.  Since we
compared \pseudox-diffusion with Laplacian or heat diffusion
(eq. \ref{minmaxDiffMean}), by default we employed Euler's method with
integration step size $\Delta t = 0.5$ and diffusion coefficient $D=1$. 
Exceptions are as follows.  \Fig[TwoCellSystem] was simulated with
$\Delta t = 0.1$ and  $\Delta t = 0.001$, respectively.  
\Fig[dneLenaPeppers], \ref{minmaxChessContrast}, and
figures \ref{dnePeppersTiles} to \ref{DynamicCompressionNoise}
were integrated with the fourth-order Runge-Kutta
method ($\Delta t = 0.5$, $D=1$).   For the compilation of \fig[Noise], again the
forth-order Runge-Kutta method was used with $\Delta t=0.01$ and $D=1/\Delta t$,
to guarantee numerical stability in the presence of high noise levels.\\
It should be emphasized that the results presented in this paper do not depend
critically on the exact value of neither $\Delta t$ and $D$, nor on the specific
choice of the integration method.  Variation of these parameters leads to a corresponding
rescaling of the time axis.  Although we exemplified the behavior of the
model only by means of two standard images which are commonly used for image
processing (\Lena and \Peppers\bs), all characteristics of the model can as
well be reproduced with other images. 
\section{\label{ProofOne}Proof of \eq[minmaxDerivative4lambda_oo_Heavi]
(for $\lambda\rightarrow\infty$ and $\lambda=0$)}
Consider the derivative of the operator  $\mathcal{T}_{\lambda}[\cdot]$
(\eq[minmaxNonlinearity]),
\formula[minmaxDerivative_T][{	\frac{\partial \mathcal{T}_{\lambda}[z]}
				{\partial z}
				= \underbrace{
					\frac{\eta}{1+\EXP}}_\mathit{term\ I}
				+ \underbrace{\frac{\eta \lambda z \EXP}
				{\left(1+\EXP \right)^2}}_\mathit{term\ II} }]
where the following three cases have to be analyzed:
\begin{description}
%
\item[Case $\lambda=0$.\ ] 
%
In this case $\eta=2$ from \eq[minmaxNormConstant], and
\formula[minmaxDerivative4lambda_0][{\frac{\partial \mathcal{T}_{0}[z]}
				{\partial z}
				= \underbrace{\frac{\eta}{1+1}}_\mathit{term\ I}
				+ \underbrace{\ \ 0\ \ }_\mathit{term\ II} = 1.}]
Thus, for $\lambda=0$ the derivative is constant one for all $z$, and
\eq[minmaxNonlinear1D] reduces to the linear diffusion \eq[minmaxLinear1D].
%
\item[Case $\lambda\rightarrow +\infty$.\ ]
%
In this case $\eta=1$ from \eq[minmaxNormConstant], and we have to consider
three additional cases according to the value of $z$.  Note that $z$ is
treated as a constant.  Hence,
\begin{description}
	\item[\underline{$z=0$.}\ ]
	\formula[minmaxDerivative4lambda_oo_zero][{	\PLIM
			\frac{\partial \mathcal{T}_{\lambda}[z]}{\partial z}
			= \underbrace{\frac{\eta}{1+1}}_\mathit{term\ I} 
			+ \underbrace{\ \ 0\ \ }_\mathit{term\ II}
			= \frac{1}{2}.}]
	This is to say that if the gradient $z$ vanishes, then the derivative
	is constant with value $1/2$.
	\item[\underline{$z>0$.}\ ]
	We start with evaluating {\it term I} of \eq[minmaxDerivative_T],
	\formula[minmaxDerivative4lambda_oo_posI][{	\PLIM
			\frac{\eta}{1+\underbrace{\PEXP}_{\rightarrow 0}}
			= \eta = 1\ \ \ \ \mathit{(term\ I)}.}]
	In the numerator of {\it term II}  appears a product of the kind
	``$\infty \cdot 0$''.  One may argue that the exponential
	$\exp(-\lambda |z|)$ always approaches zero much more faster than
	the term $\eta \lambda |z|$ is able grow (or one may equivalently
	apply l'Hospital's rule to this product by applying $d / d \lambda$
	on each factor),
	\begin{widetext}
	\formula[minmaxDerivative4lambda_oo_posII][{	\PLIM
			\frac{(\eta \lambda |z|) \cdot ( \PEXP )}
			{\left(1+\PEXP \right)^2}
			\leadsto \PLIM
			\frac{\overbrace{(\eta  |z|)}^{\mathrm{const.}}
			\cdot \overbrace{( -|z| \PEXP )}^{\rightarrow -0}}
			{(1+\underbrace{\PEXP}_{\rightarrow 0} )^2}
			= 0		\ \ \ \ \mathit{(term\ II)}.}]
	\end{widetext}
	Thus, for $\PLIM$ evaluates \eq[minmaxDerivative_T] to 1 for all $z>0$.
	\item[\underline{$z<0$.}\ ]
	Evaluating {\it term I},
	\formula[minmaxDerivative4lambda_oo_negI][{	\PLIM 
			\frac{\eta}{1+\underbrace{\NEXP}_{\rightarrow \infty}}
			= 0\ \ \ \ \mathit{(term\ I)}.}]
	Evaluating {\it term II} (again there is a little more work to do),
	\begin{widetext}
	\formula[minmaxDerivative4lambda_oo_negII][{	\PLIM
			\frac{-\eta \lambda |z|}{2 + \PEXP + \NEXP }
						\leadsto \PLIM
			\frac{\eta }{ \underbrace{\PEXP}_{\rightarrow 0}
			- \underbrace{\NEXP}_{\rightarrow \infty}}
			= 0\ \ \ \ \mathit{(term\ II)}.}]
	\end{widetext}
	Hence, for $\PLIM$ evaluates \eq[minmaxDerivative_T] to 0 for
	all $z<0$.\\

\end{description}$\Box$\\
\end{description}

Summarizing the above we saw that \eq[minmaxDerivative_T] behaves
approximately \cite{rem_heaviside} like a Heaviside function
$\heaviside$ for $\PLIM$, thus \eq[minmaxDerivative4lambda_oo_Heavi]
is proofed.  The proof of \eq[minmaxDerivative4lambda_noo_Heavi]
(for $\lambda\rightarrow\infty$) proceeds in straight analogy.


\begin{thebibliography}{41}
\expandafter\ifx\csname natexlab\endcsname\relax\def\natexlab#1{#1}\fi
\expandafter\ifx\csname bibnamefont\endcsname\relax
  \def\bibnamefont#1{#1}\fi
\expandafter\ifx\csname bibfnamefont\endcsname\relax
  \def\bibfnamefont#1{#1}\fi
\expandafter\ifx\csname citenamefont\endcsname\relax
  \def\citenamefont#1{#1}\fi
\expandafter\ifx\csname url\endcsname\relax
  \def\url#1{\texttt{#1}}\fi
\expandafter\ifx\csname urlprefix\endcsname\relax\def\urlprefix{URL }\fi
\providecommand{\bibinfo}[2]{#2}
\providecommand{\eprint}[2][]{\url{#2}}

\bibitem[{\citenamefont{Merriam-Webster}(2007)}]{Adaptation}
\bibinfo{author}{\bibnamefont{Merriam-Webster}},
  \bibinfo{journal}{Merriam-Webster Online Dictionary (16 Feb. 2007)}
  \textbf{\bibinfo{volume}{http://www.merriam-webster.com}}
  (\bibinfo{year}{2007}).

\bibitem[{\citenamefont{Walraven et~al.}(1990)\citenamefont{Walraven,
  Enroth-Cugell, Hood, MacLeod, and Schnapf}}]{WalravenEtAl90}
\bibinfo{author}{\bibfnamefont{J.}~\bibnamefont{Walraven}},
  \bibinfo{author}{\bibfnamefont{C.}~\bibnamefont{Enroth-Cugell}},
  \bibinfo{author}{\bibfnamefont{D.}~\bibnamefont{Hood}},
  \bibinfo{author}{\bibfnamefont{D.}~\bibnamefont{MacLeod}}, \bibnamefont{and}
  \bibinfo{author}{\bibfnamefont{J.}~\bibnamefont{Schnapf}}, in
  \emph{\bibinfo{booktitle}{Visual Perception: The Neurophysiological
  Foundations}}, edited by
  \bibinfo{editor}{\bibfnamefont{L.}~\bibnamefont{Spillman}} \bibnamefont{and}
  \bibinfo{editor}{\bibfnamefont{J.}~\bibnamefont{Werner}}
  (\bibinfo{publisher}{Academic Press}, \bibinfo{address}{New York},
  \bibinfo{year}{1990}), pp. \bibinfo{pages}{53--101}.

\bibitem[{\citenamefont{Carpenter and Grossberg}(1981)}]{CarpenterGrossberg}
\bibinfo{author}{\bibfnamefont{G.}~\bibnamefont{Carpenter}} \bibnamefont{and}
  \bibinfo{author}{\bibfnamefont{S.}~\bibnamefont{Grossberg}},
  \bibinfo{journal}{Journal of Theoretical Biology}
  \textbf{\bibinfo{volume}{1}}, \bibinfo{pages}{1} (\bibinfo{year}{1981}).

\bibitem[{\citenamefont{van Hateren}(2005)}]{vanHateren2005}
\bibinfo{author}{\bibfnamefont{H.}~\bibnamefont{van Hateren}},
  \bibinfo{journal}{Journal of Vision} \textbf{\bibinfo{volume}{5}},
  \bibinfo{pages}{331} (\bibinfo{year}{2005}).

\bibitem[{\citenamefont{Keil and Vitri{\`a}}(2007)}]{MatsJordi06}
\bibinfo{author}{\bibfnamefont{M.}~\bibnamefont{Keil}} \bibnamefont{and}
  \bibinfo{author}{\bibfnamefont{J.}~\bibnamefont{Vitri{\`a}}},
  \bibinfo{journal}{EURASIP Journal on Advances in Signal Processing}
  \textbf{\bibinfo{volume}{2007}}, \bibinfo{pages}{Article ID 51684, 10 pages}
  (\bibinfo{year}{2007}), \bibinfo{note}{doi:10.1155/2007/51684}.

\bibitem[{\citenamefont{Gross and Brajovic}(2003)}]{GrossBrajovic03}
\bibinfo{author}{\bibfnamefont{R.}~\bibnamefont{Gross}} \bibnamefont{and}
  \bibinfo{author}{\bibfnamefont{V.}~\bibnamefont{Brajovic}}, in
  \emph{\bibinfo{booktitle}{Springer Lecture Notes in Computer Sciences}},
  edited by \bibinfo{editor}{\bibfnamefont{J.}~\bibnamefont{Kittler}}
  \bibnamefont{and} \bibinfo{editor}{\bibfnamefont{M.}~\bibnamefont{Nixon}}
  (\bibinfo{organization}{AVBPA}, \bibinfo{year}{2003}), vol.
  \bibinfo{volume}{2688}, pp. \bibinfo{pages}{10--18}.

\bibitem[{\citenamefont{Hong and Grossberg}(2004)}]{GrossHong04}
\bibinfo{author}{\bibfnamefont{S.}~\bibnamefont{Hong}} \bibnamefont{and}
  \bibinfo{author}{\bibfnamefont{S.}~\bibnamefont{Grossberg}},
  \bibinfo{journal}{Neural Networks} \textbf{\bibinfo{volume}{17}},
  \bibinfo{pages}{787} (\bibinfo{year}{2004}).

\bibitem[{\citenamefont{Grossberg}(1983)}]{Grossberg83}
\bibinfo{author}{\bibfnamefont{S.}~\bibnamefont{Grossberg}},
  \bibinfo{journal}{Behavioral and Brain Sciences}
  \textbf{\bibinfo{volume}{6}}, \bibinfo{pages}{625} (\bibinfo{year}{1983}).

\bibitem[{\citenamefont{Carandini et~al.}(1997)\citenamefont{Carandini, Heeger,
  and Movshon}}]{CarandiniHeegerMovshon97}
\bibinfo{author}{\bibfnamefont{M.}~\bibnamefont{Carandini}},
  \bibinfo{author}{\bibfnamefont{D.}~\bibnamefont{Heeger}}, \bibnamefont{and}
  \bibinfo{author}{\bibfnamefont{J.}~\bibnamefont{Movshon}},
  \bibinfo{journal}{The Journal of Neuroscience} \textbf{\bibinfo{volume}{17}},
  \bibinfo{pages}{8621} (\bibinfo{year}{1997}).

\bibitem[{\citenamefont{Carandini and Heeger}(1994)}]{CarHee94}
\bibinfo{author}{\bibfnamefont{M.}~\bibnamefont{Carandini}} \bibnamefont{and}
  \bibinfo{author}{\bibfnamefont{D.}~\bibnamefont{Heeger}},
  \bibinfo{journal}{Science} \textbf{\bibinfo{volume}{264}},
  \bibinfo{pages}{1333} (\bibinfo{year}{1994}).

\bibitem[{\citenamefont{Heeger}(1992)}]{Heeger92}
\bibinfo{author}{\bibfnamefont{D.}~\bibnamefont{Heeger}},
  \bibinfo{journal}{Visual Neuroscience} \textbf{\bibinfo{volume}{9}},
  \bibinfo{pages}{181} (\bibinfo{year}{1992}).

\bibitem[{rem({\natexlab{a}})}]{rem_rowscols}
\bibinfo{note}{Moving along rows corresponds to discrete $y$-values, and moving
  along columns corresponds to discrete values of $x$}.

\bibitem[{rem({\natexlab{b}})}]{rem_adiabatic}
\bibinfo{note}{Adiabatic boundary conditions are implemented in the simulations
  by reducing the diffusion operators at the domain borders to existing
  neighbors}.

\bibitem[{\citenamefont{Koenderink}(1984)}]{Koenderink84}
\bibinfo{author}{\bibfnamefont{J.}~\bibnamefont{Koenderink}},
  \bibinfo{journal}{Biological Cybernetics} \textbf{\bibinfo{volume}{50}},
  \bibinfo{pages}{363} (\bibinfo{year}{1984}).

\bibitem[{\citenamefont{Lifshitz and Pizer}(1990)}]{LifshitzPizer1990}
\bibinfo{author}{\bibfnamefont{L.}~\bibnamefont{Lifshitz}} \bibnamefont{and}
  \bibinfo{author}{\bibfnamefont{S.}~\bibnamefont{Pizer}},
  \bibinfo{journal}{IEEE Transactions on Pattern Analysis and Machine
  Intelligence} \textbf{\bibinfo{volume}{12}}, \bibinfo{pages}{529}
  (\bibinfo{year}{1990}).

\bibitem[{\citenamefont{Jobson et~al.}(1997)\citenamefont{Jobson, Rahmann, and
  Woodell}}]{JobsonRahmanWoodell97}
\bibinfo{author}{\bibfnamefont{D.}~\bibnamefont{Jobson}},
  \bibinfo{author}{\bibfnamefont{Z.}~\bibnamefont{Rahmann}}, \bibnamefont{and}
  \bibinfo{author}{\bibfnamefont{G.}~\bibnamefont{Woodell}},
  \bibinfo{journal}{IEEE Transactions on Image Processing}
  \textbf{\bibinfo{volume}{6}}, \bibinfo{pages}{965} (\bibinfo{year}{1997}).

\bibitem[{\citenamefont{Shannon}(1948)}]{Shannon48}
\bibinfo{author}{\bibfnamefont{C.}~\bibnamefont{Shannon}},
  \bibinfo{journal}{Bell Systems Technical Journal}
  \textbf{\bibinfo{volume}{27}}, \bibinfo{pages}{379} (\bibinfo{year}{1948}).

\bibitem[{\citenamefont{Wainwright}(1999)}]{Wainwright99}
\bibinfo{author}{\bibfnamefont{M.}~\bibnamefont{Wainwright}},
  \bibinfo{journal}{Vision Research} \textbf{\bibinfo{volume}{39}},
  \bibinfo{pages}{3960} (\bibinfo{year}{1999}).

\bibitem[{\citenamefont{Raviola and Gilula}(1975)}]{RaviolaGilula75}
\bibinfo{author}{\bibfnamefont{E.}~\bibnamefont{Raviola}} \bibnamefont{and}
  \bibinfo{author}{\bibfnamefont{N.~B.} \bibnamefont{Gilula}},
  \bibinfo{journal}{Journal of Cell Biology} \textbf{\bibinfo{volume}{65}},
  \bibinfo{pages}{192} (\bibinfo{year}{1975}).

\bibitem[{\citenamefont{Kolb}(1977)}]{Kolb77}
\bibinfo{author}{\bibfnamefont{H.}~\bibnamefont{Kolb}},
  \bibinfo{journal}{Journal of Neurocytology} \textbf{\bibinfo{volume}{6}},
  \bibinfo{pages}{131} (\bibinfo{year}{1977}).

\bibitem[{\citenamefont{Nelson et~al.}(1985)\citenamefont{Nelson, Lynn,
  Dickinson-Nelson, and Kolb}}]{NelsonEtAl85}
\bibinfo{author}{\bibfnamefont{R.}~\bibnamefont{Nelson}},
  \bibinfo{author}{\bibfnamefont{T.}~\bibnamefont{Lynn}},
  \bibinfo{author}{\bibfnamefont{A.}~\bibnamefont{Dickinson-Nelson}},
  \bibnamefont{and} \bibinfo{author}{\bibfnamefont{H.}~\bibnamefont{Kolb}}, in
  \emph{\bibinfo{booktitle}{Neurocircuitry of the Retina: {A} Cajal Memorial}}
  (\bibinfo{publisher}{Elsevier}, \bibinfo{address}{New York},
  \bibinfo{year}{1985}), pp. \bibinfo{pages}{109--121}.

\bibitem[{\citenamefont{Mills and Massey}(1994)}]{MillsMassey94}
\bibinfo{author}{\bibfnamefont{S.~L.} \bibnamefont{Mills}} \bibnamefont{and}
  \bibinfo{author}{\bibfnamefont{S.~C.} \bibnamefont{Massey}},
  \bibinfo{journal}{Visual Neuroscience} \textbf{\bibinfo{volume}{11}},
  \bibinfo{pages}{549} (\bibinfo{year}{1994}).

\bibitem[{\citenamefont{Mills and Massey}(2000)}]{MillsMassey00}
\bibinfo{author}{\bibfnamefont{S.~L.} \bibnamefont{Mills}} \bibnamefont{and}
  \bibinfo{author}{\bibfnamefont{S.~C.} \bibnamefont{Massey}},
  \bibinfo{journal}{Journal of Neuroscience} \textbf{\bibinfo{volume}{20}},
  \bibinfo{pages}{8629} (\bibinfo{year}{2000}).

\bibitem[{\citenamefont{Gibson et~al.}(1999)\citenamefont{Gibson, Beierlein,
  and Connors}}]{GiBeCo99}
\bibinfo{author}{\bibfnamefont{J.}~\bibnamefont{Gibson}},
  \bibinfo{author}{\bibfnamefont{M.}~\bibnamefont{Beierlein}},
  \bibnamefont{and} \bibinfo{author}{\bibfnamefont{B.}~\bibnamefont{Connors}},
  \bibinfo{journal}{Nature} \textbf{\bibinfo{volume}{402}}, \bibinfo{pages}{75}
  (\bibinfo{year}{1999}).

\bibitem[{\citenamefont{Galarreta and Hestrin}(1999)}]{GalHes99}
\bibinfo{author}{\bibfnamefont{M.}~\bibnamefont{Galarreta}} \bibnamefont{and}
  \bibinfo{author}{\bibfnamefont{S.}~\bibnamefont{Hestrin}},
  \bibinfo{journal}{Nature} \textbf{\bibinfo{volume}{402}}, \bibinfo{pages}{72}
  (\bibinfo{year}{1999}).

\bibitem[{\citenamefont{Galarreta and Hestrin}(2002)}]{GalHes02}
\bibinfo{author}{\bibfnamefont{M.}~\bibnamefont{Galarreta}} \bibnamefont{and}
  \bibinfo{author}{\bibfnamefont{S.}~\bibnamefont{Hestrin}},
  \bibinfo{journal}{Proceedings of the National Academy of Sciences USA}
  \textbf{\bibinfo{volume}{99}}, \bibinfo{pages}{12438} (\bibinfo{year}{2002}).

\bibitem[{\citenamefont{Naka and Rushton}(1967)}]{NakaRushton67}
\bibinfo{author}{\bibfnamefont{K.-I.} \bibnamefont{Naka}} \bibnamefont{and}
  \bibinfo{author}{\bibfnamefont{W.}~\bibnamefont{Rushton}},
  \bibinfo{journal}{Journal of Physiology} \textbf{\bibinfo{volume}{193}},
  \bibinfo{pages}{437} (\bibinfo{year}{1967}).

\bibitem[{\citenamefont{Lamb}(1976)}]{Lamb76}
\bibinfo{author}{\bibfnamefont{T.}~\bibnamefont{Lamb}},
  \bibinfo{journal}{Journal of Physiology} \textbf{\bibinfo{volume}{263}},
  \bibinfo{pages}{239} (\bibinfo{year}{1976}).

\bibitem[{\citenamefont{Winfree}(1995)}]{Winfree95}
\bibinfo{author}{\bibfnamefont{A.}~\bibnamefont{Winfree}}, in
  \emph{\bibinfo{booktitle}{The Handbook of Brain Theory and Neural Networks}},
  edited by \bibinfo{editor}{\bibfnamefont{M.}~\bibnamefont{Arbib}}
  (\bibinfo{publisher}{The MIT Press}, \bibinfo{address}{Cambridge,
  Massachusetts}, \bibinfo{year}{1995}), pp. \bibinfo{pages}{1054--1056}.

\bibitem[{\citenamefont{Benda et~al.}(2001)\citenamefont{Benda, Bock, Rujan,
  and Ammerm{\"u}ller}}]{BendaEtAl01}
\bibinfo{author}{\bibfnamefont{J.}~\bibnamefont{Benda}},
  \bibinfo{author}{\bibfnamefont{R.}~\bibnamefont{Bock}},
  \bibinfo{author}{\bibfnamefont{P.}~\bibnamefont{Rujan}}, \bibnamefont{and}
  \bibinfo{author}{\bibfnamefont{J.}~\bibnamefont{Ammerm{\"u}ller}},
  \bibinfo{journal}{Visual Neuroscience} \textbf{\bibinfo{volume}{18}},
  \bibinfo{pages}{835} (\bibinfo{year}{2001}).

\bibitem[{\citenamefont{Edwards et~al.}(1991)\citenamefont{Edwards, Heitler,
  Leise, and Friscke}}]{Edwards91}
\bibinfo{author}{\bibfnamefont{D.}~\bibnamefont{Edwards}},
  \bibinfo{author}{\bibfnamefont{W.}~\bibnamefont{Heitler}},
  \bibinfo{author}{\bibfnamefont{E.}~\bibnamefont{Leise}}, \bibnamefont{and}
  \bibinfo{author}{\bibfnamefont{R.}~\bibnamefont{Friscke}},
  \bibinfo{journal}{Journal of Neuroscience} \textbf{\bibinfo{volume}{11}},
  \bibinfo{pages}{2117} (\bibinfo{year}{1991}).

\bibitem[{\citenamefont{Edwards et~al.}(1998)\citenamefont{Edwards, Yeh, and
  Krasne}}]{Edwards98}
\bibinfo{author}{\bibfnamefont{D.}~\bibnamefont{Edwards}},
  \bibinfo{author}{\bibfnamefont{S.-R.} \bibnamefont{Yeh}}, \bibnamefont{and}
  \bibinfo{author}{\bibfnamefont{F.}~\bibnamefont{Krasne}},
  \bibinfo{journal}{Proceedings of the National Academy of Sciences USA}
  \textbf{\bibinfo{volume}{9}}, \bibinfo{pages}{7145} (\bibinfo{year}{1998}).

\bibitem[{\citenamefont{Bukauskas et~al.}(2002)\citenamefont{Bukauskas, Angele,
  Verselis, and Bennett}}]{BukauskasEtAl02}
\bibinfo{author}{\bibfnamefont{F.}~\bibnamefont{Bukauskas}},
  \bibinfo{author}{\bibfnamefont{A.}~\bibnamefont{Angele}},
  \bibinfo{author}{\bibfnamefont{V.}~\bibnamefont{Verselis}}, \bibnamefont{and}
  \bibinfo{author}{\bibfnamefont{M.}~\bibnamefont{Bennett}},
  \bibinfo{journal}{Proceedings of the National Academy of Sciences USA}
  \textbf{\bibinfo{volume}{99}}, \bibinfo{pages}{7113} (\bibinfo{year}{2002}).

\bibitem[{\citenamefont{Piccolino et~al.}(1984)\citenamefont{Piccolino, Neyton,
  and Gerschenfeld}}]{PiccolinoEtAl84}
\bibinfo{author}{\bibfnamefont{M.}~\bibnamefont{Piccolino}},
  \bibinfo{author}{\bibfnamefont{J.}~\bibnamefont{Neyton}}, \bibnamefont{and}
  \bibinfo{author}{\bibfnamefont{H.}~\bibnamefont{Gerschenfeld}},
  \bibinfo{journal}{The Journal of Neuroscience} \textbf{\bibinfo{volume}{4}},
  \bibinfo{pages}{2477} (\bibinfo{year}{1984}).

\bibitem[{\citenamefont{Keil et~al.}(2005)\citenamefont{Keil, Crist{\'o}bal,
  Hansen, and Neumann}}]{MatsEtAl05}
\bibinfo{author}{\bibfnamefont{M.}~\bibnamefont{Keil}},
  \bibinfo{author}{\bibfnamefont{G.}~\bibnamefont{Crist{\'o}bal}},
  \bibinfo{author}{\bibfnamefont{T.}~\bibnamefont{Hansen}}, \bibnamefont{and}
  \bibinfo{author}{\bibfnamefont{H.}~\bibnamefont{Neumann}},
  \bibinfo{journal}{Neural Networks} \textbf{\bibinfo{volume}{18}},
  \bibinfo{pages}{1319} (\bibinfo{year}{2005}).

\bibitem[{\citenamefont{Yu et~al.}(2002)\citenamefont{Yu, Giese, and
  Poggio}}]{YuGiesePoggio02}
\bibinfo{author}{\bibfnamefont{A.}~\bibnamefont{Yu}},
  \bibinfo{author}{\bibfnamefont{M.}~\bibnamefont{Giese}}, \bibnamefont{and}
  \bibinfo{author}{\bibfnamefont{T.}~\bibnamefont{Poggio}},
  \bibinfo{journal}{Neural Computation} \textbf{\bibinfo{volume}{14}},
  \bibinfo{pages}{2857} (\bibinfo{year}{2002}).

\bibitem[{\citenamefont{Keil}(2003)}]{MatsThesis}
\bibinfo{author}{\bibfnamefont{M.}~\bibnamefont{Keil}}, Ph.D. thesis,
  \bibinfo{school}{Universit{\"a}t Ulm, Faculty for Computer Science},
  \bibinfo{address}{Ulm, Germany} (\bibinfo{year}{2003}).

\bibitem[{\citenamefont{Riesenhuber and Poggio}(1999)}]{RiesenhuberPoggio99}
\bibinfo{author}{\bibfnamefont{M.}~\bibnamefont{Riesenhuber}} \bibnamefont{and}
  \bibinfo{author}{\bibfnamefont{T.}~\bibnamefont{Poggio}},
  \bibinfo{journal}{Nature Neuroscience} \textbf{\bibinfo{volume}{2}},
  \bibinfo{pages}{1019} (\bibinfo{year}{1999}).

\bibitem[{\citenamefont{Riesenhuber and Poggio}(2000)}]{RiesenhuberPoggio00}
\bibinfo{author}{\bibfnamefont{M.}~\bibnamefont{Riesenhuber}} \bibnamefont{and}
  \bibinfo{author}{\bibfnamefont{T.}~\bibnamefont{Poggio}},
  \bibinfo{journal}{Nature Neuroscience} \textbf{\bibinfo{volume}{3}},
  \bibinfo{pages}{1199} (\bibinfo{year}{2000}).

\bibitem[{\citenamefont{Press et~al.}(1997)\citenamefont{Press, Teukolsky,
  Vetterling, and Flannery}}]{Recipes}
\bibinfo{author}{\bibfnamefont{W.}~\bibnamefont{Press}},
  \bibinfo{author}{\bibfnamefont{S.}~\bibnamefont{Teukolsky}},
  \bibinfo{author}{\bibfnamefont{W.}~\bibnamefont{Vetterling}},
  \bibnamefont{and} \bibinfo{author}{\bibfnamefont{B.}~\bibnamefont{Flannery}},
  \emph{\bibinfo{title}{Numerical Recipes in {C} - The Art of Scientific
  Computing (2nd Edition)}} (\bibinfo{publisher}{Cambridge University Press},
  \bibinfo{address}{http://www.ulib.org/webRoot/Books/Numerical{\_}Recipes/},
  \bibinfo{year}{1997}).

\bibitem[{rem({\natexlab{c}})}]{rem_heaviside}
\bibinfo{note}{Not exactly, because of \eq[minmaxDerivative4lambda_oo_zero]}.

\end{thebibliography}
\end{document}